\documentclass[useAMS,usenatbib]{mn2e}
\usepackage{graphicx,subfigure,ram}

\begin{document}

\title[Low Surface Brightness Galaxies]{Studying the radio continuum from 
nuclear activity and star formation in  Giant Low Surface Brightness Galaxies}
\author[Mishra et. al.]{Alka Mishra $^{1}$\thanks{E-mail:
alkam7@gmail.com}
N.~G.~Kantharia$^{2}$ M.~Das$^{3}$ D.~C.~Srivastava $^1$ S.~N.~Vogel $^{4}$\\
$^{1}$Department of Physics, D.D.U. Gorakhpur University,Gorakhpur,India\\
$^{2}$National Centre for Radio Astrophysics,TIFR ,Pune, India\\
$^{3}$Indian Institute of Astrophysics, Koramangala,Bangalore,India\\
$^{4}$Department of Astronomy, University of Maryland, College Park, MD 20742, USA}


\maketitle

\begin{abstract}

We present a multifrequency radio continuum study of seven 
giant low surface brightness (GLSB) galaxies using the Giant 
Metrewave Radio Telescope (GMRT).  
GLSB galaxies are optically faint, dark-matter dominated systems
that are poorly evolved and have large H{\sc i} gas disks.
Our sample consists of 
GLSB galaxies that show signatures of nuclear activity in their optical spectra. 
We detect radio emission from the nuclei of all the seven galaxies. 
Five galaxies have nuclear spectral indices that range from 0.12 to -0.44  
and appear to be core-dominated; the two galaxies have  a 
steeper spectrum.  Two of the galaxies, UGC 2936 and UGC 4422 
show significant radio emission from  their disks.  
In our 610 MHz observations of UGC 6614, we detect radio lobes associated 
with the radio-loud active galactic nucleus (AGN). The lobes have 
a spectral index of -1.06$\pm$0.12. 
The star formation rates estimated from the radio emission, for the entire sample
range from 0.15 to 3.6 M$_{\odot}$ yr$^{-1}$. 
We compare the radio images with the near-ultraviolet 
(NUV) images from GALEX and near-infrared (NIR) images from 2MASS. 
The galaxies present a diversity of relative NUV, NIR and 
radio emission,  supporting an episodic star formation scenario 
for these galaxies. 
Four galaxies are classified members of groups and one is classified as isolated. 
Our multiwavlength study of this sample suggests that the environment plays 
an important role in the evolution of these galaxies.
\end{abstract}

\begin{keywords}
{galaxies: individual-  UGC 1378, UGC 1922,  UGC 2936, 
UGC 4422,  Malin 2, UGC 6614, UM 163}
\end{keywords}

\section{Introduction}
Low surface brightness (LSB) galaxies  are late type spirals 
that have properties quite distinct from regular spirals on the 
Hubble sequence.  They are characterized by a central disk 
surface brightness fainter than 23 mag arcsec$^{-2}$ in the 
B band (\citealt{impey97}; \citealt{geller12}). They have diffuse 
stellar disks \citep{deblok95}, large H{\sc i} disks and
massive dark matter halos (\citealt{mcgaugh98}; 
\citealt{vanden00}). The low star formation rates \citep{oneil07} 
and low metallicities \citep{mcgaugh94} of these galaxies 
suggest that they are less evolved compared to high surface 
brightness galaxies (HSBGs). Chemical evolution models
indicate that star formation has proceeded in a stochastic manner 
in LSB galaxies and hence their evolution 
is slow (\citealt{bell00}; 
\citealt{vanden00}).

Studies show that LSB galaxies span a wide 
range of morphologies, from the more populous dwarf 
LSB  galaxies \citep{pustilnik11} to the relatively 
larger giant spirals like Malin-1 (\citealt{mcgaugh95}; 
\citealt{beijersbergen99}; \citealt{galaz02}). The larger 
LSB  galaxies
or the so-called giants are relatively rare (\citealt{sprayberry95}; 
\citealt{bothun90}) and often isolated compared to HSBGs 
\citep{galaz11}. In fact, they are often found close to the edge 
of voids \citep{rosenbaum09}. Although the disks of giant 
low surface brightness (GLSB) galaxies are usually poor in 
star formation \citep{auld06}; weak but distinct 
spiral arms are often present \citep{mcgaugh95}. The barred galaxy
fraction is low compared to HSBGs \citep{mayer04} and a significant 
fraction is bulgeless. The lack of disk evolution may be due to the  
massive dark matter halos that dominate the disks of these galaxies 
as halos play an important role in slowing down the formation of disk 
instabilities \citep{mihos97}.  However, there are exceptions; some 
GLSB galaxies show vigorous signs of disk star formation and emit copious  
UV and H$\alpha$ emission \citep{boissier08}.

\begin{table*}
 \centering
  \begin{minipage}{150mm}
   \caption{Properties of the Sample Galaxies}
\begin{tabular}{@{}llllllllll@{} }\hline
\footnotetext[1]{NASA Extragalactic Database} 
\footnotetext[2]{Hyperleda} 
\footnotetext[3]{Lyon Group of Galaxies (LGG) \citep{garcia93}}
\footnotetext[4]{Low Density Contrast Extended (LDCE) groups (\citealt{crook07};\citealt{crook08})}

Galaxies & Type$^{a}$&$\alpha_{(2000)}^{a}$ & $\delta_{(2000)}^{a}$&Optical size$^{a}$&inclination$^{b}$&V$_{hel}^{a}$
     & 1$^{\prime\prime}$$^{a}$ &  Environment & Recent  \\

&&    (h:m:s)&($\circ$ : $^\prime$ : $^{\prime\prime}$) &( $^\prime$ $\times$ $^\prime$ ) & (degree) &
                 (km s$^{-1}$) & =(kpc) &   &  Supernova\\ \hline

UGC 1378 & (R)SB(rs)a& 01:56:19.2&+73:16:58 & 3.4 $\times$ 2.3 & 68.4 &  2935 &0.20& .... & .... \\
UGC 1922 &    S      & 02:27:45.9&+28:12:32 & 2.1 $\times$ 1.6 & 33.9 & 10894 &0.70& LDCE 0163(37)$^{d}$& 1989s(Type Ia)\\
UGC 2936 & SB(s)d    & 04:02:48.2&+01:57:58 & 2.5 $\times$ 0.7 & 80.6 &  3813 &0.24& .... & 1991bd(Type II)\\
UGC 4422 & SAB(rs)c  & 08:27:42.0&+21:28:45 & 1.6 $\times$ 1.3 & 48.8 &  4330 &0.30& LGG 159$^{c}$ &1999aa(Type Ia) \\
         &           &           &          &                  &      &       &    & LDCE 571(22)$^{d}$ \\
Malin-2  &   Sd/p    & 10:39:52.5&+20:50:49 & 1.5 $\times$ 0.9 & 35.6 & 13830 &0.94&  Isolated & .... \\
UGC 6614 & (R)SA(r)a & 11:39:14.9&+17:08:37 & 1.7 $\times$ 1.4 & 29.8 &  6352 &0.45& LDCE 829(3)$^{d}$& ....\\
UM 163   & SB(r)b pec& 23:30:32.3&-02:27:45 & 1.9 $\times$ 1.3 & 65.0 & 10022 &0.68& LDCE 1583(3)$^{d}$ & ....\\
\hline
\end{tabular}
\end{minipage}
\label{tab:table.1}
\end{table*}

GLSB galaxies have low surface brightness disks and some have 
been found to host active 
galactic nuclei (AGN) (\citealt{sprayberry95}; \citealt{schombert98}). 
\cite{schombert98} studied a sample of GLSB galaxies using optical spectroscopy 
and concluded that at least 30\% of the sample showed AGN activity, 
especially those that have prominent bulges.  AGN in GLSB galaxies have also 
been detected in X-ray  (\citealt{das09}; \citealt{naik10}) 
with similar luminosities as of  AGN in bright spirals 
($10^{40} -10^{42}~$erg~s$^{-1}$). In their optical study of stellar 
populations in bulges of low surface brightness galaxies, \citet{morelli12} 
find them to be similar to HSBGs which they say implies that the disk-bulge
evolution is decoupled.
 
LSB galaxies have not been extensively 
investigated in the radio continuum.
Recently, \cite{mei09} have 
conducted a survey of AGNs in 196 LSB galaxies
sample by  \cite{impey96} using the spectroscopic 
data of SDSS DR5 and the corresponding FIRST data.  They find that 
about $10-20 \%$  GLSB galaxies host an AGN as compared to the 
50\% found for HSBGs (\citealt{kennicutt89a}; \citealt{ho97}).
\cite{boissier08} have examined the UV colours from GALEX
and the star formation efficiency (SFE) of these galaxies.  
They find that the UV light extends out farther than the optical 
light and SFE are lower for some LSB galaxies.  
They also find that the UV emission of a few LSB galaxies in the sample resembles that
observed in the XUV-disk galaxies (\citealt{gildepaz05};  \citealt{thilker07}).
Besides, they also find that the  FUV-NUV colour for LSB galaxies  
is redder than expected for star forming galaxies which they interpret 
as being due to bursts of star formation.  In this paper we present a detailed
multiwavelength (either 1280 or 1420 MHz (L band), 610~MHz, 325 and 240~MHz) radio 
study of a sample of seven GLSB galaxies with the Giant Metrewave 
Radio Telescope (GMRT).  We study the radio morphology of the 
sources,  the  star forming component and the spectrum of 
the central active source in these galaxies.
We refer to AGNs as a flat-spectrum sources if $\alpha$ $\geq$ -0.5. 
Synchrotron self-absorption results in such flat spectrum.
We also present a comparative study of the radio, UV (NUV) 
and NIR (J-band) emissions from these galaxies.

\begin{figure*}
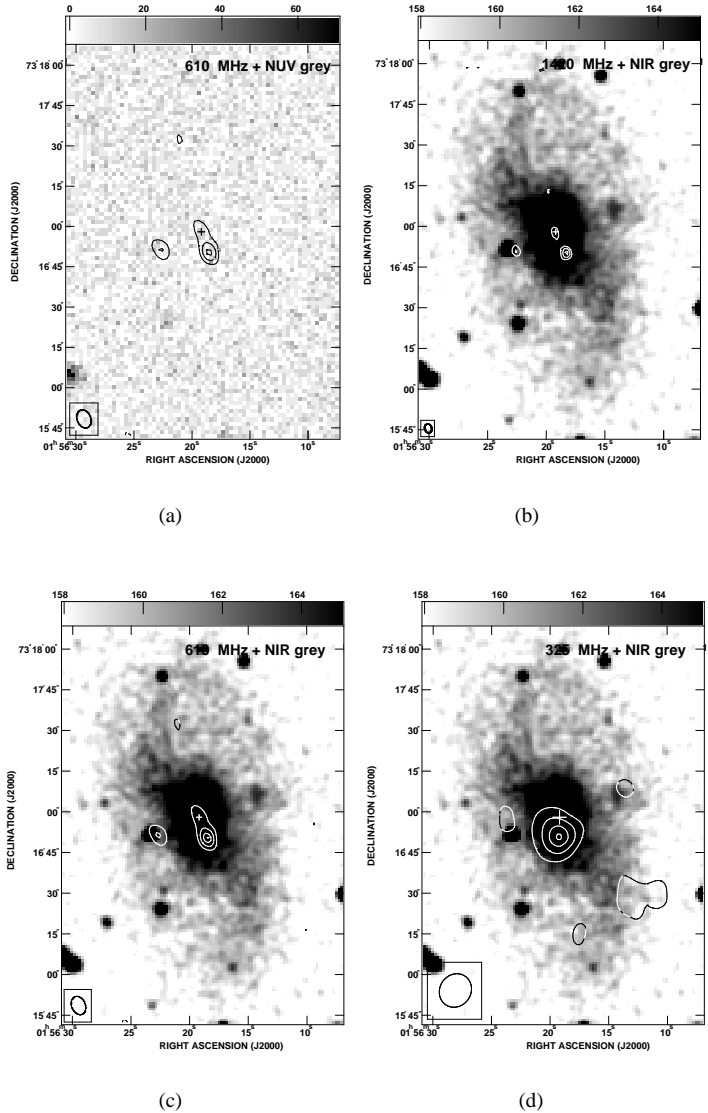

\subfigure[]{\includegraphics[height=7.0cm]{figure.1a.PS}}
\subfigure[]{\includegraphics[height=7.0cm]{figure.1b.PS}}\\
\subfigure[]{\includegraphics[height=7.0cm]{figure.1c.PS}}
\subfigure[]{\includegraphics[height=7.0cm]{figure.1d.PS}}\\
\caption{\textbf{Images of UGC 1378}:   The cross marks the position of the optical centre.
(a)  The contours showing the 610 MHz emission are plotted at 60 $\times$ 
(-8,-4,4,8,12,14) $\mu$Jy beam$^{-1}$. The angular resolution is 7$^{\prime\prime}$ 
$\times$ 5$^{\prime\prime}$, PA = 21$^{\circ}$.82.  
The NUV grey scale is  counts/sec.
(b) The contour levels of the 1420 MHz emission are 28$\times$ (-8,-4,4,8,13) 
$\mu$Jy beam$^{-1}$ for a  beamsize of 3$^{\prime\prime}$ $\times$ 2$^{\prime\prime}$,
PA = 16$^{\circ}$.26. 
The NIR grey scale are data-number.
(c) Contour levels are similar to (a).  
(d) The contours showing the 325 MHz emission are plotted
at  0.2  $\times$ (-6,-4,4,6,8,9.3) mJy beam$^{-1}$. The beamsize is 
13$^{\prime\prime}$ $\times$ 11$^{\prime\prime}$, PA = -33$^{\circ}$.39.}
\label{fig:figure.1}
\end{figure*}

\section {The Sample}
\footnotetext[1]{See http://irsa.ipac.caltech.edu/applications/2MASS/IM/}
\footnotetext[2]{See http://galex.stsci.edu/GR6/?page=tilelist$\&$survey=allsurveys}
In this paper, we present the radio continuum study of a sample of 
seven GLSB galaxies with an aim to study the AGN. 
The sample is selected from several studies in the literature 
(\citealt{schombert98}; \citealt{sprayberry95}; \citealt{ramya11}) 
 based on the following criteria:
(1) large spirals  (2)  prominent bulge
(3) presence of an optically identified AGN and 
(4) v$_{sys}$ $\leq$ 15,000 km $s^{-1}$.  
The sample consists of UGC 1378, UGC 1922, UGC 2936, UGC 4422, 
Malin 2, UGC 6614 and UM 163.  
The basic properties of the  galaxies are listed in
Table \ref{tab:table.1}.  
The galaxy morphologies  range from  Sa
to Sd types.  Four galaxies in our sample are classified as members
of groups.
Supernovae explosions have been recorded in three galaxies;
UGC 1922 (SN 1989s) \citep{mueller89}, UGC 2936 (SN 1991bd) \citep{mueller91}
and UGC 4422 (SN 1999aa) \citep{armstrong99} in
the recent past. 
The optical identifications of AGNs are generally done using 
the spectral features of [NII], [SII], [OIII], H$\alpha$, H$\beta$ and 
several other emission lines.

\begin{table}
\begin{minipage}{150mm}
   \caption{Details of GALEX images used in the paper}
    \begin{tabular}{@{}llllll@{}}\hline
\footnotetext[1]{All sky Imaging Survey (AIS), Nearby Galaxy Survey (NGS)}
\footnotetext {Medium Imaging Survey (MIS), Guest Investigator  Program (GIP) }
\footnotetext[2]{Taken from \citet{martin05}}
Galaxy  &  Survey$^{a}$ & NUV$_{exposure time}$$^{b}$ &  Depth$^{b}$ \\
        &               &         (ksec)                &   (M$_{AB}$)       \\ \hline

UGC~1378 &  AIS & 0.1     &  20.5 \\
UGC~1922 &  NGS & 1.5     &  23 \\
UGC~2936 &  GIP & 3.3     &  .... \\
UGC~4422 &  MIS & 1.5     &  23 \\
MALIN-2  &  GIP & 7.6     & .... \\
UGC~6614 &  NGS & 1.5     &  23  \\
UM~163   &  GIP & 3.3     & ....  \\
\hline
\end{tabular}
\end{minipage}
\label{tab:table.2}
\end{table}

Although LSB galaxies do not 
contain abundant quantities of molecular gas, CO observations have 
revealed the presence of molecular gas in UGC~1922
\citep{oneil03} and in UGC~6614, Malin~2 (\citealt{das06}; \citealt{das10}). 
Four galaxies in our sample namely UGC~1378, UGC~1922, UGC~2936 and UGC ~6614
have been observed in X-rays \citep{das09}.  While a compact nuclear source has 
been detected in X-rays from UGC~2936 \citep{das09} and UGC ~6614 \citep{naik10},  
diffuse X-ray emission is detected in UGC 1378, UGC 1922 and UGC 2936 \citep{das09}.  
Millimetre wave radio continuum has been detected from the core in 
UGC~6614 at 111 GHz \citep{das06}.   All the sample 
galaxies have been observed in the NIR (2MASS, J band) and in the UV (GALEX, NUV)
and detected in the radio continuum at 1.4 GHz (NVSS, 1.4 GHz). 
We have used  2MASS$^{1}$ J band images  (wavelength=1.25 microns) 
which has a magnitude limit of 15.0 for extended sources and a magnitude limit of
15.8 for point sources in our analysis.   The NIR bands trace old stellar 
populations  \citep{skrutskie06} so these images are used to understand the same.
We have used the near-UV (NUV; 1770-2730$\AA$)  images downloaded 
from the GALEX website$^{2}$  as tracer of young massive 
stellar regions \citep{martin05}.   Table \ref{tab:table.2} gives the
details of the GALEX data we have used.  
UGC 1378 which is observed as  part of the  All sky Imaging Survey (AIS) of GALEX, 
has a short exposure time of  100 s while
rest of the galaxies have longer exposure times ranging from $\sim$  1500 to 7500 s.
The cores of all the galaxies and
the star forming disks of UGC 1378, UGC 2936 and UGC 4422 are detected 
in the NIR.

\section{Observations And Data Analysis}
\footnotetext[3]{AIPS (Astronomical Image Processing System) is 
distributed by the National Radio Astronomy Observatory 
(NRAO), which is a facility of 
the National Science Foundation operated under cooperative
agreement by Associated Universities, Inc.}

The observations were done using the GMRT \citep{swarup91}. 
The GMRT is an interferometric 
array of thirty antennas, each antenna being 45~m in diameter.
Observations were carried out from  December 2005 to October 2012 
at 240 MHz, 325 MHz, 610 MHz and  L band.  
Flux calibration was done using scans on one of the standard 
calibrators 3C~147, 3C~286 and 3C~48, 
which were observed at the start and end of the observing runs. 
Phase calibration was done using VLA calibrator sources observed 
before and after each scan on the target source. Details of the 
observations are given in Table \ref{tab:table.3}. 
The raw data were converted to FITS and imported to AIPS$^{3}$. 
The data were reduced using standard tasks in AIPS. 
The calibrator data were edited and gain solutions obtained.
The flux density calibrators were used for band-pass calibration. 
The band-pass calibrated data were averaged   over a smaller  number of 
channels to avoid bandwidth smearing effects.
Wide-field imaging techniques were applied to the data.  We 
generated 49 facets at 240 MHz, 25 facets at 325, 25 facets at 
610 MHz and 9 facets at  L band  across the primary
beam \citep{cornwell92}. Data sets were phase self-calibrated up to three iterations by 
using strong point sources within the primary beam field. A final 
round of phase and amplitude self-calibration was then done. Naturally (Robust =5)  
and uniformly  (Robust =0) weighted  \citep{briggs95} images at different resolutions were
made at all the frequencies and then corrected for primary beam 
attenuation. In order to study the compact nuclear emission  we also made images after 
excluding  the  data corresponding to large angular scales in the image i.e. short baselines.

\begin{table*}
 \centering
  \begin{minipage}{130mm}
   \caption{Details of GMRT Observations}
    \begin{tabular}{@{}rrrllllllllrrrr@{}}\hline

\footnotetext[1]{Flux densities (S$_{phase}$) of phase calibrators from GETJY, task in AIPS}
\footnotetext[2]{Flux densities (S$_{flux}$) of flux calibrators from SETJY, task in AIPS}
\footnotetext[3]{Observing Bandwidth}
\footnotetext[4]{Average number of antennas working during the observations}

Galaxies& Obs. Date &  Band & \multicolumn{6}{c}{Calibrators}&&  BW$^{c}$& t$_{source}$ & N$_{Ant.}$ $^{d}$  \\
        &(dd/mm/yy) & (MHz)  &&   Phase &  S$_{phase}$(Jy)$^{a}$ &&  Flux &  S$_{flux}$ (Jy)$^{b}$ &   & (MHz) & (hrs)   \\
\cline{4-9}  \\ \hline

UGC 1378&26/10/12 & 325  && 0114+483 &  5.9  &&3C48        &  45.0         && 32   & 3.5 &  28 \\
        &30/07/06 & 610  && 0217+738 &  2.1  &&3C48, 3C147 &  29.4, 38.2   && 16   & 6.0 &  27 \\
        &30/12/11 &1420  && 0217+738 &  1.8  &&3C48        &  16.8         && 16  & 5.5 &  29 \\
UGC 1922&27/10/12 & 325  && 3C48     & 44.5  &&3C48        &  44.5         && 32   &  3.5 &  28 \\
        &31/12/05 & 610  && 3C48     & 29.3  &&3C48        &  29.3         && 16   &  6.0 &  28 \\
        &31/12/11 & 1420 && 3C48     & 16.4  &&3C48        &  16.4         && 16  &  6.0 &  29 \\
UGC 2936&14/05/07 & 610  && 0323+055 &  5.3  &&3C147       & 38.2          && 16   & 3.0 &  28 \\
        &19/05/07 & 1280 && 0323+055 &  3.4  &&3C147       & 23.6          && 16  &  3.0 &  28 \\
UGC 4422&26/06/11 & 325  && 0909+428 & 17.6  &&3C147, 3C286&  54.0,26.4    && 32   & 2.0 &  28 \\
        &14/05/07 & 610  && 0735+331 &  5.3  &&3C147       &  38.2         && 16   & 2.5 &  27 \\
        &29/12/11 & 1420 && 0842+185 &  1.1  &&3C286, 3C147& 22.2, 14.8    && 16  &  5.5 &  29 \\
Malin-2 &09/01/06 & 240  && 3C241    & 10.1  &&3C147, 3C286&  48.6, 25.8   && 16   & 5.5 &  28 \\
        &09/01/06 & 610  && 3C241    &  4.3  &&3C147, 3C286& 36.4, 20.6    && 16   & 5.5 &  28 \\
        &11/06/11 & 1280 && 3C241    &  2.0  &&3C147       &  23.7         && 32  & 1.0 &  27 \\
UGC 6614& 31/12/05& 240  && 1123+055 &  6.4  &&3C48, 3C286 & 47.3, 25.8    && 16   & 6.5 &  27 \\
        & 31/12/05& 610  && 1123+055 &  3.4  &&3C48, 3C286 &  29.3, 20.7   && 16   & 6.5 &  27 \\
        & 12/07/11& 1280 && 1120+143 &  2.7  &&3C147       & 23.5          && 32  & 1.45&  29 \\
UM 163 &26/06/11  & 325  && 2225-049 & 13.9  &&3C48, 3C147 &  42.8, 46.7   && 32   & 2.5 &  29 \\
       &01/07/06  & 610  && 2225-049 & 10.6  &&3C48        &   29.4        && 16   &   4.0& 28 \\
       &28/07/06  & 1420 && 2225-049 &  8.4  &&3C48, 3C286 & 16.2,15.0     && 04  & 5.0 &  27 \\
\hline
\end{tabular}
\end{minipage}
\label{tab:table.3}
\end{table*}

\section{Results}

We detect radio emission from 
the nuclear region of all the seven galaxies at one or more
radio bands.  Two of the 
galaxies, UGC~2936 and  UGC~4422, show diffuse emission  
which appears to be 
associated with the disk. 
The results for the images made with Robust =5 and
Robust=0  \citep{briggs95} are summarized  in 
Table \ref{tab:table.4}.

\begin{figure*}
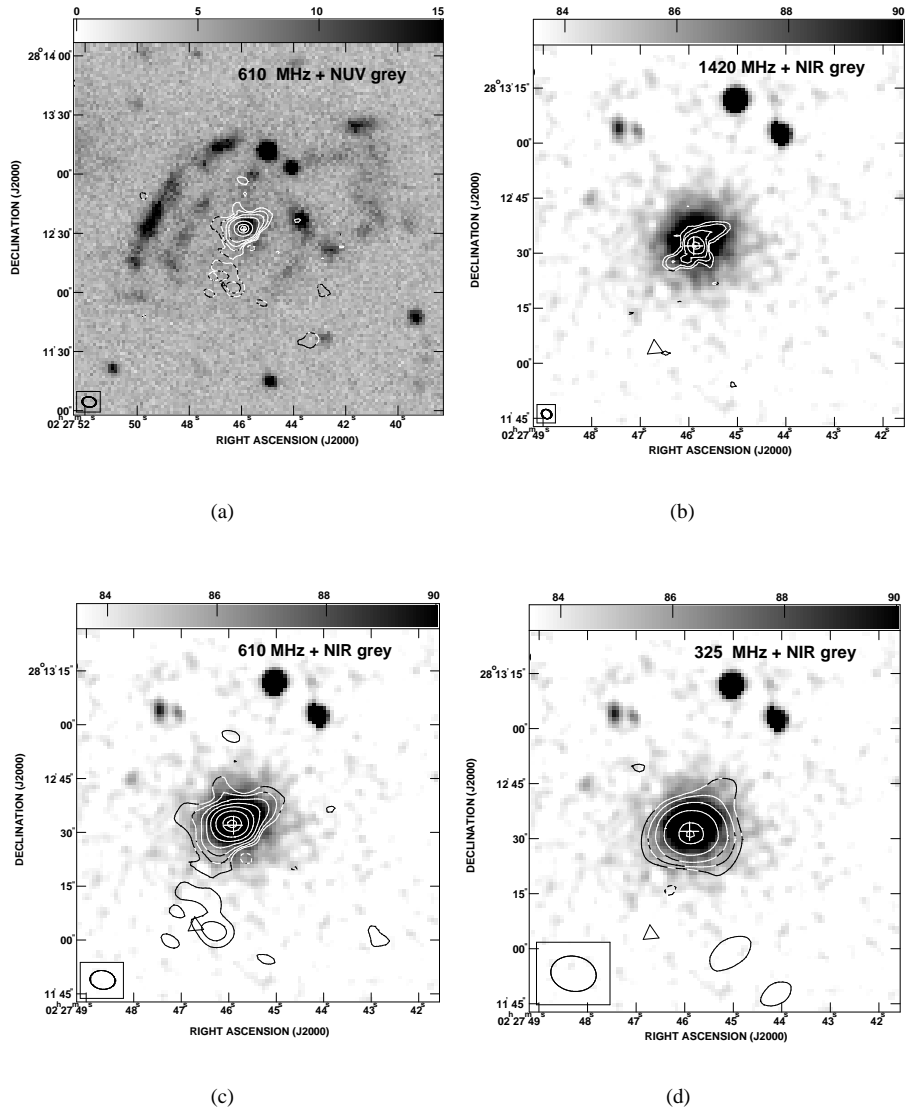

\subfigure[]{\includegraphics[height=7.0cm]{figure.2a.PS}}
\subfigure[]{\includegraphics[height=7.0cm]{figure.2b.PS}}\\
\subfigure[]{\includegraphics[height=7.0cm]{figure.2c.PS}}
\subfigure[]{\includegraphics[height=7.0cm]{figure.2d.PS}}\\
\caption{\textbf{Images of UGC 1922}: The cross marks the position of optical centre and 
the triangle  at the south-east marks the position of the supernova, SN 1989S.
(a) The contours showing the 610 MHz emission are plotted at 0.1 $\times$
(-8,-4,4,8,16,32,64,128,256,330) mJy beam$^{-1}$. The angular resolution  is
7$^{\prime\prime}$ $\times$ 5$^{\prime\prime}$, PA= 82$^{\circ}$.11.
The NUV grey scale is  counts/sec.
(b) The contour levels of the 1420 MHz emission are 60 $\times$ 
(-8,-4,4,8,16,64,256,350) $\mu$Jy beam$^{-1}$ for beamsize of  
3$^{\prime\prime}$ $\times$ 2$^{\prime\prime}$, PA = 54$^{\circ}$.55.
The NIR grey scale are data-number.
(c) Contour levels are similar to (a).
(d) The contours showing the 325 MHz emission are plotted at  
0.8 $\times$ (-8,-4,4,8,16,32,64,75) mJy beam$^{-1}$. 
The beamsize is 12$^{\prime\prime}$ $\times$ 9$^{\prime\prime}$, PA = 79$^{\circ}$.68.}
\label{fig:figure.2}
\end{figure*}

We present the images in Figures \ref{fig:figure.1}- \ref{fig:figure.8} 
using the following format. 
In the first panel the 610 MHz radio continuum contours are 
overlaid on the grey scale GALEX NUV images of the galaxies. 
In the other zoomed-in  panels, the contours of radio emission 
are overlaid  on the NIR 2MASS  images. 
In Table \ref{tab:table.5}, we list the spectral indices 
for the nuclear emission and the integrated emission.
The nuclear emission is estimated from images made after excluding shorter 
baselines  which trace extended emission.
In this table we have listed (i)  Column 3 : nuclear flux densities at
GMRT bands after convolving the images to the lowest resolution.
(ii) Column 5 :  the estimated  total flux densities at GMRT
bands after convolving the images to the lowest resolution
(iii) Column 7 : NVSS flux densities and (iv) Column 8 : the differences
between the NVSS flux and the nuclear emission estimated from GMRT high 
resolution images at L band.
In the following subsections  
we  discuss the results on the  individual galaxies.
The  spectra are shown in Figures  \ref{fig:figure.9} (a)- (f).

\begin{table*}
 \centering
  \begin{minipage}{200mm}
   \caption{Results of GMRT  Observations}
   \begin{tabular}{@{}rrllllllllllllr@{}}\hline 

Galaxy&Band &\multicolumn{5}{c}{ Low resolution (Robust=5)$^{a}$}&&\multicolumn{4}{c}
{High resolution (Robust=0)$^{a}$} & Structure\\ 
\cline{4-7}  \cline{9-12} 
\footnotetext[1]{ See \citet{briggs95}}
\footnotetext[2] {A systematic flux density error of 15\% of total flux at 240, 325 
                  and 610 MHz and 5\% at  L band, is assumed.}
\footnotetext[3] {Galaxy emission is fragmented so the flux density is not measured. }
\footnotetext[4] {UGC~6614 is not detected at 240 MHz and 4$\sigma$ limit is tabulated.}

    &       && Flux Density$^{b}$  & Beam Size & P.A. & $\sigma$ && Flux Density$^{b}$ & Beam Size& P.A. & $\sigma$ &  \\
& MHz && mJy &($^{\prime\prime} \times^{\prime\prime}$)&  degree & mJy/b  &&  mJy& ($^{\prime\prime} \times ^{\prime\prime}$)
&  degree & mJy/b  \\ \hline

UGC 1378&  325 &&     2.2$\pm$0.3   &32 $\times$ 18&  60.4  & 0.43 &&    1.7$\pm$0.2  & 13 $\times$ 11 & -33.3 & 0.18 & Compact core \\
        &  610 &&     1.7$\pm$0.2   &11 $\times$ 7 &   8.0  & 0.08 &&    0.9$\pm$0.1  &  7 $\times$ 5  &  21.8 & 0.07 & Compact core \\
        & 1420 &&    0.92$\pm$0.04  & 5 $\times$ 4 &   7.1  & 0.04 &&    0.54$\pm$0.02 &  3 $\times$ 2 &  16.2 & 0.03 & Compact core \\
UGC 1922& 325  &&    72.8$\pm$10.9  &31 $\times$ 16 &  57.2  & 1.44 &&    68.1$\pm$10.2 & 12 $\times$ 9 &  79.6 & 0.80 & Compact core    \\
        & 610  &&    64.5$\pm$9.6   &15 $\times$ 12& -82.8  & 0.20 &&    56.0$\pm$8.4  &  7 $\times$ 5 &  82.1 & 0.11 & Compact core+extended \\
        & 1420 &&    35.1$\pm$1.7   & 4 $\times$ 3 &  51.9  & 0.13 &&    33.4$\pm$1.6  &  3 $\times$ 2 &  54.5 & 0.06 & Compact core    \\
UGC 2936& 610  &&    48.0$\pm$7.2   &20 $\times$ 14& -42.3  & 0.28 &&    46.4$\pm$6.9  & 15 $\times$ 5 &  89.4 & 0.34 & Compact core+extended \\
        & 1280 &&    29.3$\pm$1.4   & 6 $\times$ 4 & -21.8  & 0.08 &&    ....$^{c}$   &   3 $\times$ 2 &  41.4 & 0.08 & Compact core+extended \\
UGC 4422&  325 &&    49.9$\pm$7.4   &26 $\times$ 18& -11.6 & 0.52 &&    42.8$\pm$6.4  & 10 $\times$ 9  &  44.5 & 0.30 & Compact core+extended\\
        &  610 &&    10.1$\pm$1.5   &18 $\times$ 4 & -26.7 & 0.10 &&     7.5$\pm$1.1  &  6 $\times$ 4  &  -35.4 & 0.12 & Compact core \\
        & 1420 &&     4.0$\pm$0.2   & 5 $\times$ 2 &  59.3 & 0.03 &&     3.6$\pm$0.1   & 3 $\times$ 2  &   61.0 & 0.02 & Compact core  \\
Malin-2 &  240 &&     5.2$\pm$0.7   &45 $\times$ 30& -23.3 & 0.76 &&     5.2$\pm$0.7  & 37 $\times$ 14 &  -31.9 & 0.94 & Compact core \\
        &  610 &&     5.0$\pm$0.7   &20 $\times$ 19&  33.5 & 0.18 &&     5.0$\pm$0.7  &  7 $\times$ 5  &   82.6 & 0.29 & Compact core \\
        & 1280 &&     3.6$\pm$0.1   &9  $\times$ 6 & -89.1 & 0.09 &&     3.6$\pm$0.1  &  3 $\times$ 2  &  -76.8 &0.07 & Compact core \\
UGC 6614&  240 &&   $<$6.8$^{d}$    &40 $\times$ 36&  -4.6 & 1.71 &&     $<$4.6$^{d}$ &16 $\times$ 14  & -64.7 &1.15 & Not detected \\
        &  610 &&    10.8$\pm$1.6   &20 $\times$ 16& -50.2 & 0.16 &&    10.2$\pm$1.5  &  8 $\times$ 6  & -85.9 &0.15& Compact core+extended\\
        & 1280 &&     4.4$\pm$0.2   &7  $\times$  4& 10.5  & 0.04 &&     4.4$\pm$0.2  &  3 $\times$ 2 & 74.2 & 0.05 & Compact core \\
UM 163  &  325 &&    27.9$\pm$4.1   &17 $\times$ 11& 61.8 & 0.27 &&    26.1$\pm$3.9  & 12 $\times$ 9  & 64.7 & 0.25 & Compact core \\
        &  610 &&    14.8$\pm$2.2   &16 $\times$ 10&-8.36 & 0.23 &&    13.5$\pm$2.0  &  7 $\times$ 5 & 59.4 &0.19 & Compact core+extended \\
        & 1420 &&     8.8$\pm$0.4   &12 $\times$ 5 & -28.9 & 0.14 &&     7.1$\pm$0.3  &  5 $\times$ 3  &  36.0& 0.17 & Compact core \\
\hline
\end{tabular}
\end{minipage}
\label{tab:table.4}
\end{table*}

\subsection{Results on individual galaxies}

\textbf{UGC 1378}: \cite{schombert98} identified the optical 
AGN in this galaxy.  Diffuse X-ray emission is detected 
from the central parts of this galaxy \citep{das09}.  
The picture of environment is not clear for this galaxy.

This galaxy is bright in the NIR band with emission arising 
from the entire disk (Figures \ref{fig:figure.1} (b)-(d)) 
but no NUV emission is detected  which is likely due to the short exposure 
time (Table \ref{tab:table.2}).  
The  compact radio emission from the galaxy, which is  
detected at 325 MHz, gets resolved into two peaks 
with $\sim$ 8$^{\prime\prime}$ offset  at 1420 and 610 MHz. 
In fact, the radio peak emission at all the three frequency bands 
is detected to the  south of the optical centre of the galaxy whereas 
fainter emission is detected from 
the optical AGN at 1420 and 610 MHz. 
We convolved the images of the galaxy at 610 and 1420 MHz to 
the lower resolution of the 325 MHz map to estimate the 
spectrum of the emission.

We detect a compact source to the east of the 
centre of the galaxy at the right ascension 01$^{h}$56$^{m}$22.6$^{s}$  
and declination +73$^{\circ}$16$^{\prime}$51$^{\prime\prime}$.5 
at 610 and 1420~MHz (Figures \ref{fig:figure.1} (b) and (c)).  
This is close to a compact NIR source which is
likely part of the galaxy.  This source is not detected in our 325 MHz map nor 
is it visible in the NVSS map.
The estimated  fluxes for the eastern source at 610 and 1420 MHz 
are 0.52$\pm$0.07  mJy  and  0.25$\pm$0.01  mJy which gives a spectral index of
$\alpha$ $\sim$  -0.9$\pm$0.1. No radio  emission from the star forming 
disk is detected.

\textbf{UGC 1922}: 
AGN activity has been observed in the optical spectrum of the 
galaxy by \cite{schombert98}.  
The galaxy hosted a type Ia supernova,
1989S \citep{mueller89}.   The galaxy has been detected in CO
emission and the molecular gas 
($\sim$ 1.1 $\times$ 10$^{9}$ M$_{\odot}$) appears to be concentrated 
within the inner  30$^{\prime\prime}$ of the disk \citep{oneil03}. 
Diffuse X-ray emission has also been detected from the central parts of this 
galaxy \citep{das09}. This galaxy is classified as a member of the 
37 member-group LDCE 0163  \citep{crook08}. 

Radio emission is detected from  the central parts of the galaxy 
(Figure \ref{fig:figure.2} (a)). 
Bright NUV emission is detected along the spiral arm situated 
in the north-east of the centre of
the galaxy and from other compact regions. 
The central parts of the galaxy are bright in radio emission
and NIR  (Figures \ref{fig:figure.2} (b)-(d)).  
The 610 MHz image shows emission from the central part 
with an extension towards the south,  
close to the reported SN 1989S. The 1420 MHz image resolves the 
emission in the central part of this low inclination 
galaxy into a mini-spiral in the centre with a bright peak.  
Interestingly two massive star forming complexes located north-west to the 
centre of the galaxy are detected in the NUV and NIR.  No radio
continuum emission is detected in any of the observed radio bands.

\textbf{UGC 2936}: This galaxy is almost edge-on in morphology.
The AGN in this GLSB galaxy was identified by \citet{sprayberry95}. 
The AGN  is detected in X-ray  with a  luminosity of 
1.8 $\times$ 10$^{42}$ ergs s$^{-1}$ \citep{das09}. 
A type II supernova, SN 1991bd has been recorded in this galaxy 
\citep{mueller91}.
Unlike most GLSB galaxies with bulges, it has a boxy/peanut shaped
bulge rather than a classical bulge.  This suggests that the galaxy
may be undergoing secular evolution of its bar into a boxy bulge
\citep{raha91}.

\begin{figure*}
\subfigure[]{\includegraphics[height=7cm]{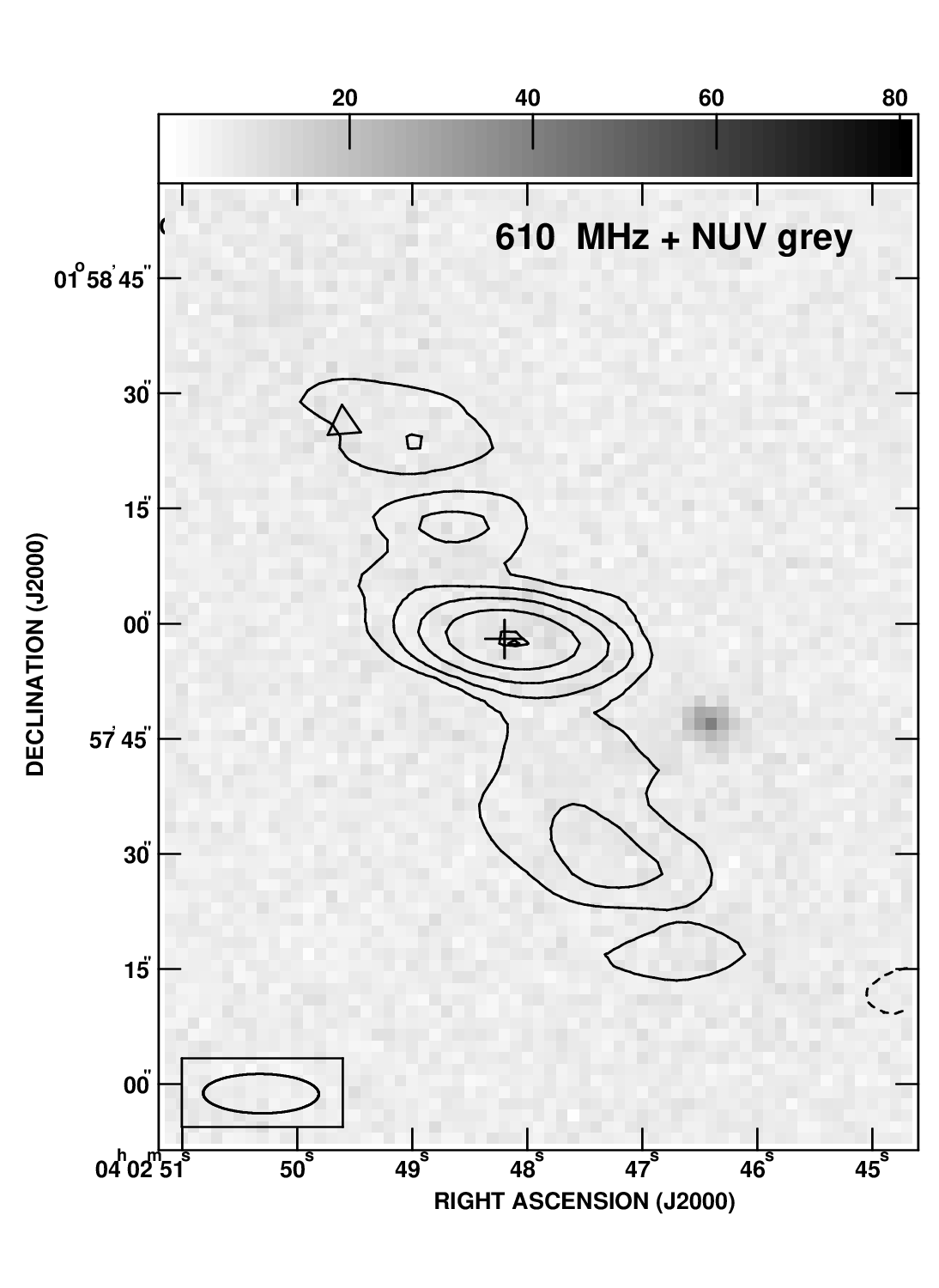}}
\subfigure[]{\includegraphics[height=7cm]{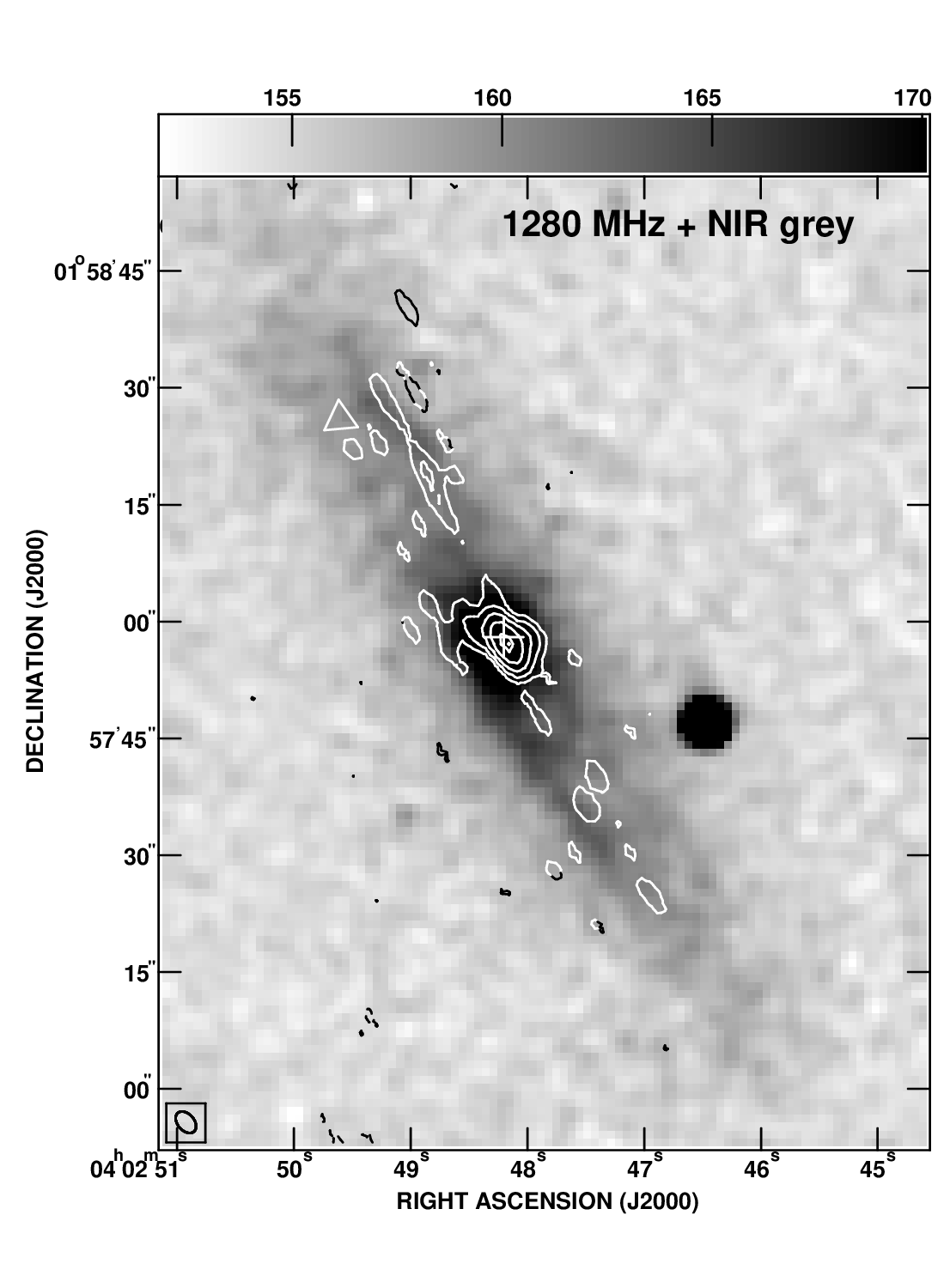}}
\subfigure[]{\includegraphics[height=7cm]{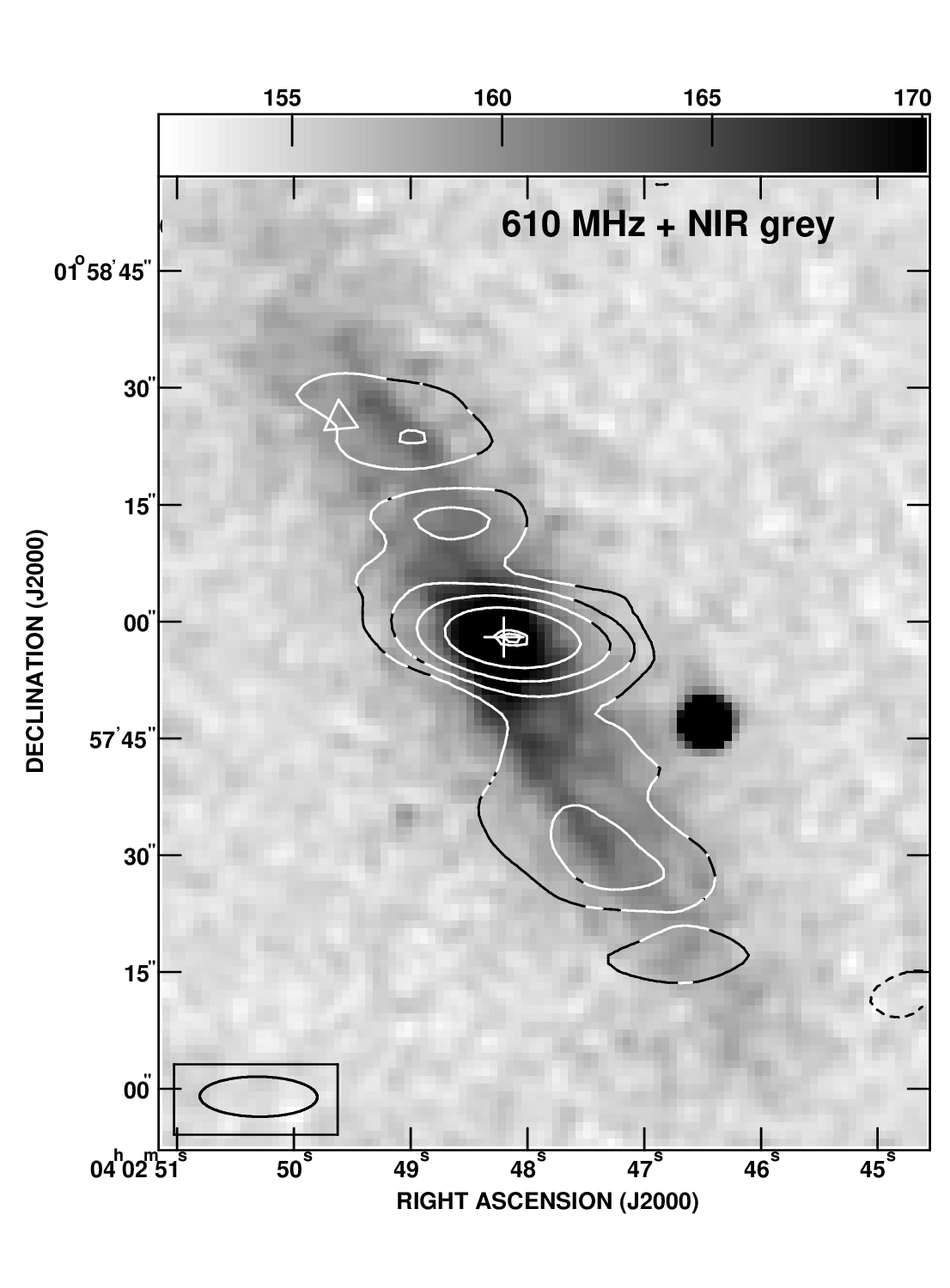}}
\caption{\textbf{Images of UGC 2936}: The cross  marks the optical centre of the galaxy and 
the triangle in the north-east marks the position of the supernova, SN 1991bd.
(a) The contours showing the 610 MHz emission are plotted at  
0.3 $\times$ (-8,-4,4,8,16,32,64,66) mJy beam$^{-1}$. The angular resolution  
is 15$^{\prime\prime}$ $\times$ 5$^{\prime\prime}$, PA = 89$^{\circ}$.48.
The NUV grey scale is  counts/sec.
(b) The contour levels of the 1280 MHz emission are  
76 $\times$ (-8,-4,4,8,16,32,64,75) $\mu$Jy beam$^{-1}$
for beamsize of 3$^{\prime\prime}$ $\times$ 2$^{\prime\prime}$, PA =41$^{\circ}$.44.
The NIR grey scale are data-number.
(c) Contour  levels are similar to (a).}
\label{fig:figure.3}
\end{figure*}

The galaxy is observed at 1280 and 610 MHz with the GMRT. 
Faint NUV  emission is detectable from the centre of the 
galaxy (Figure \ref{fig:figure.3} (a)).  
The entire disk of the galaxy is detected 
in NIR and radio bands in addition to the intense emission 
from the active nucleus  (Figures \ref{fig:figure.3}
(b) and (c)). 
Good correlation is seen between the extent of the NIR and 
radio disks due to star formation.

\textbf{UGC 4422}: 
The AGN in this barred galaxy was identified by \cite{schombert98}.  
No X-ray emission has been detected from the centre of this galaxy 
with an upper limit of $ 10^{39}$ erg-s$^{-1}$ \citep{das09}.
This galaxy hosted a type Ia supernova SN 1999aa \citep{armstrong99}.
It is a member of the group LGG 159 \citep{garcia93} 
and LDCE 571 with 22 members \citep{crook08}.

\begin{figure*}
\subfigure[]{\includegraphics[height=7.0cm]{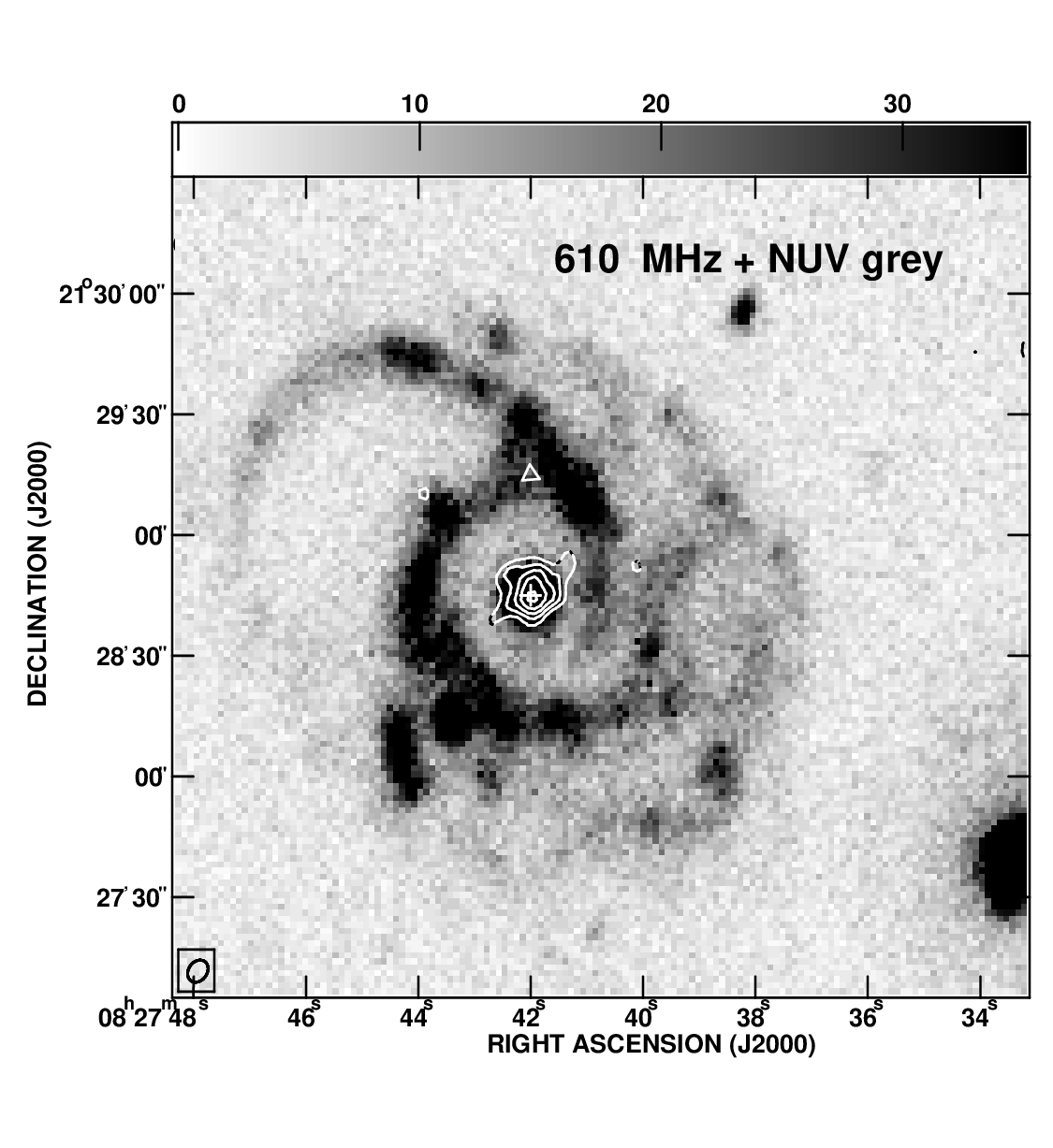}}
\subfigure[]{\includegraphics[height=7.0cm]{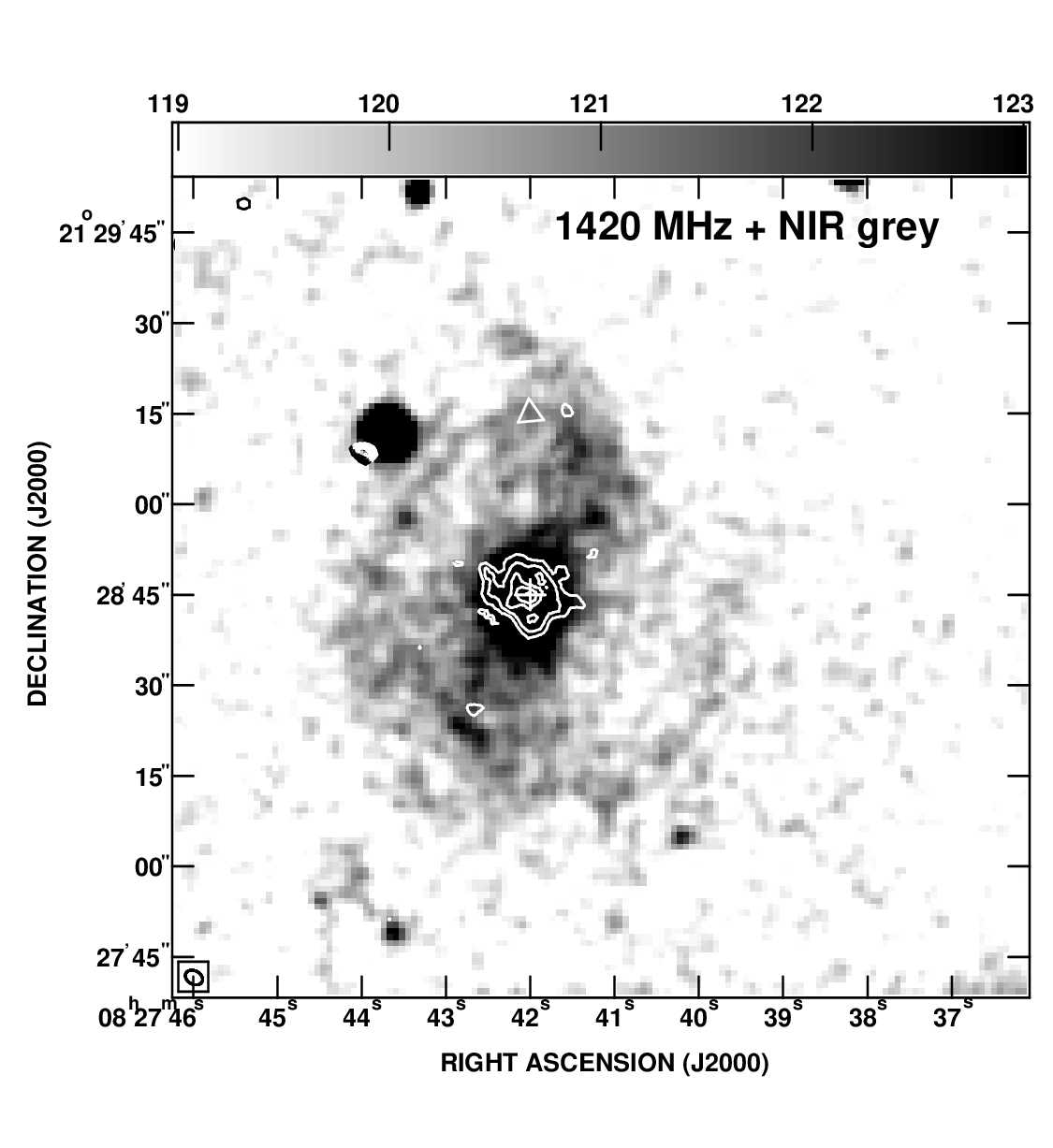}}\\
\subfigure[]{\includegraphics[height=7.0cm]{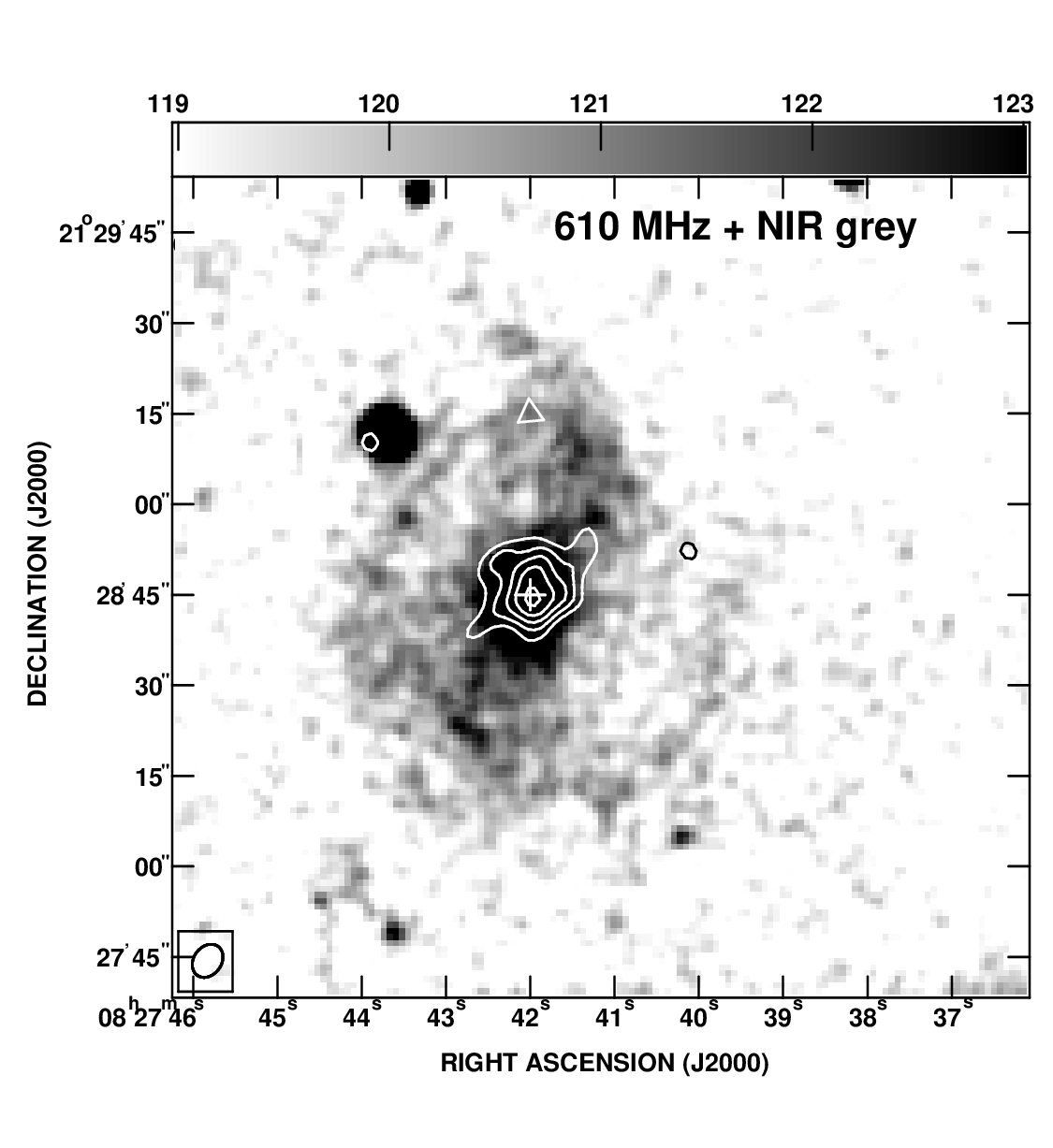}}
\subfigure[]{\includegraphics[height=7.0cm]{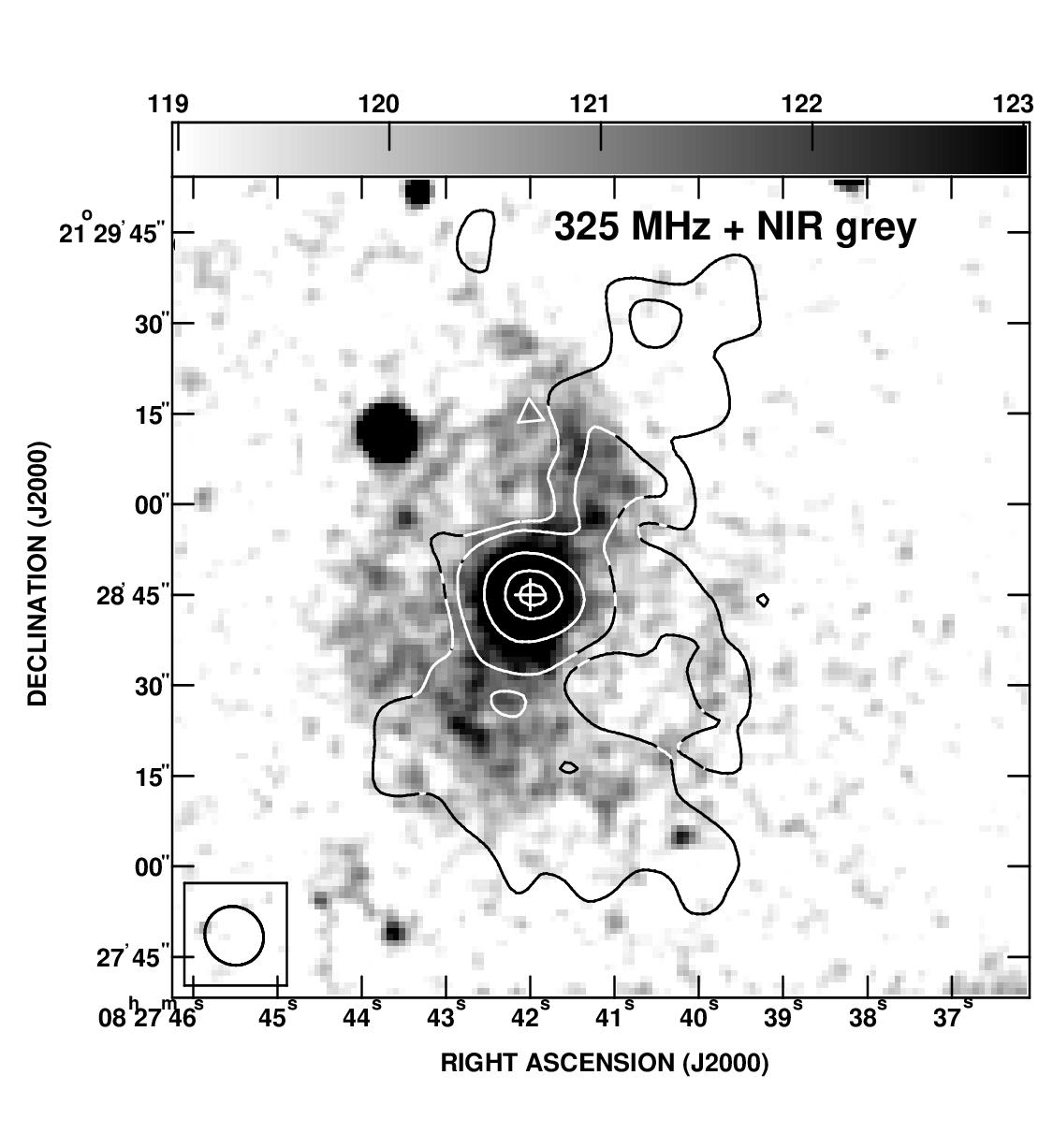}}\\
\caption{\textbf{Images of UGC 4422}: The cross marks the optical centre of the galaxy 
and triangle at the north marks the position of the supernova, SN 1999aa.
(a)  The contours showing the 610 MHz emission are plotted at  0.1 $\times$
(-6,-4,4,6,8,10,12) mJy beam$^{-1}$. The angular  resolution is 6$^{\prime\prime}$ 
$\times$ 4$^{\prime\prime}$, PA = -35$^{\circ}$.49.
The NUV grey scale is  counts/sec.
(b) The contour levels of the 1420 MHz emission are 28 $\times$
(-6,-4,4,6,8,10,13) $\mu$Jy beam$^{-1}$ for beamsize  3$^{\prime\prime}$ 
$\times$ 2$^{\prime\prime}$, PA = 61$^{\circ}$.05. The NIR grey scale are data-number.
(c) Contour  levels  are similar to (a).
(d) The contours showing the 325 MHz emission are plotted at  0.4 $\times$ 
(-4,-3,3,4,8,10,12,14) mJy beam$^{-1}$. The beamsize is   
10$^{\prime\prime}$ $\times$ 9$^{\prime\prime}$, PA = 44$^{\circ}$.53.}
\label{fig:figure.4}
\end{figure*}

The 610 MHz emission is confined to the central parts of the 
galaxy while bright NUV emission is observed along the spiral arms 
and to the west of the centre are also seen 
(Figure  \ref{fig:figure.4} (a)).
The emission at 1420~MHz and 610~MHz is concentrated in the 
central parts of the galaxy while  the 325~MHz emission  
extends along the spiral arms (Figures  \ref{fig:figure.4} (b)- (d)).  
The 610 MHz image  shows emission at the onset of the spiral arms in 
the north-west and south-east directions.  Interestingly, the NIR emission from
this galaxy is extended.
The diffuse extended emission seen at 325 MHz along the spiral arms 
is not detected at 610 or 1420 MHz.
The diffuse emission at 325 MHz in the southern arm has
a brightness of 1.1 mJy/beam (see Figure \ref{fig:figure.4} (d)).  
Using the $4 \sigma$ limit of our low resolution map at 610 MHz of 0.48 mJy/beam;
we estimate that the spectral index is $<-1.3$ for the diffuse emission.
We made maps after excluding the 
shorter baselines  to remove  extended emission  but did not  detect
an unresolved source in any of the wavebands.

\begin{figure}
{\includegraphics[height=7.0cm]{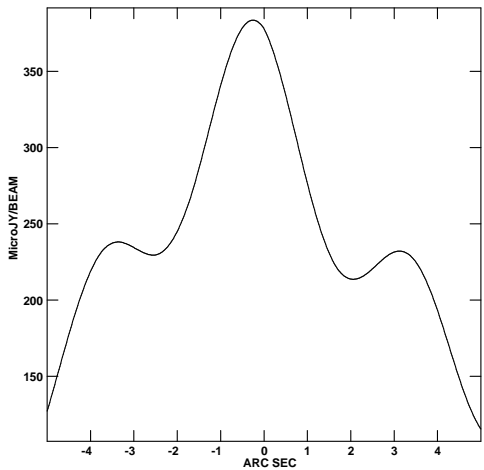}}\\
\caption{A slice along the right ascension passing through the
centre of the galaxy in the 1420 MHz image of UGC 4422 which
shows the presence of a central peak and a circumnuclear ring.}
\label{fig:figure.5}
\end{figure}

The optical image of this galaxy shows a circumnuclear ring-like feature at the centre
which is also discernible in our 1420 MHz image.  To confirm  its
presence in our 1420 MHz image, we took a one-dimensional cut along  right ascension
(see Figure \ref{fig:figure.5}).  This shows the presence of three peaks; the 
outer ones likely define a ring surrounding the central peak which is likely the optical AGN. 
We measure the size of the ring $\sim$ 6$^{\prime\prime}$ which at the distance
of UGC 4422 corresponds to a diameter of 1.8 kpc.  \cite{comeron10}   detect 
a star forming circumnuclear ring in their HST optical image of the galaxy. 
They estimate the  semimajor axis of the ring to be 5.4$^{\prime\prime}$ which corresponds to
1.6 kpc.  
A type Ia supernova, SN 1999aa was reported in the northern part of the galaxy 
but we do not detect any radio emission from that region.

\textbf{Malin 2}:  
The  nucleus of the galaxy shows AGN  activity at optical wavelengths  
(e.g. \citealt{ramya11}). It has been detected
in CO emission  and the molecular gas ($\sim$ 2.6 $\times$ 10$^{9}$ M$_{\odot}$) 
is found to extend towards the west in the galaxy disk (\citealt{das06}; \citealt{das10}).  
It is not detected in the continuum at millimetre wavelengths \citep{das06}.  
This galaxy is classified as an isolated galaxy.

\begin{figure*}
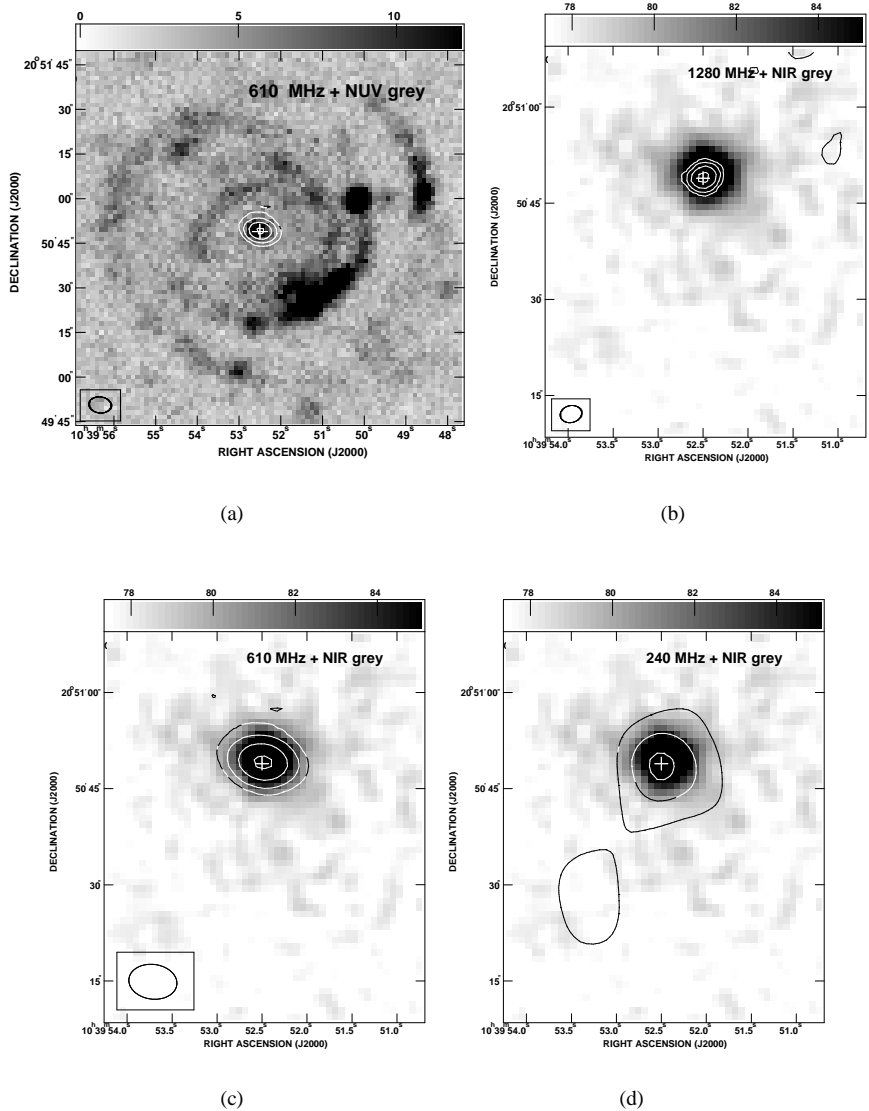

\subfigure[]{\includegraphics[height=7.0cm]{figure.6a.PS}}
\subfigure[]{\includegraphics[height=7.0cm]{figure.6b.PS}}\\
\subfigure[]{\includegraphics[height=7.0cm]{figure.6c.PS}}
\subfigure[]{\includegraphics[height=7.0cm]{figure.6d.PS}}\\
\caption{\textbf{Images of Malin 2}: The cross marks the optical centre of the galaxy.
(a) The contours showing the 610 MHz emission are plotted at 
0.2 $\times$ (-8,-4,4,8,16,28) mJy beam$^{-1}$.   
The angular resolution  is 7$^{\prime\prime}$ $\times$ 5$^{\prime\prime}$, 
PA =82$^{\circ}$.69.  
The NUV grey scale is  counts/sec.
(b) The contour levels of the 1280 MHz emission are 89 $\times$  
(-8,-4,4,8,16,32)  $\mu$Jy beam$^{-1}$ for  beamsize is 3$^{\prime\prime}$ 
$\times$ 2$^{\prime\prime}$, PA =-76$^{\circ}$.85. The NIR grey scale are data-number.
(c) Contour levels are similar to (a). 
(d) The contours showing the 240 MHz emission are plotted at   0.8 $\times$
(-6,-4,4,6,7.5) mJy beam$^{-1}$. The  beamsize is 37$^{\prime\prime}$ 
$\times$ 14$^{\prime\prime}$, PA = -31$^{\circ}$.92. }
\label{fig:figure.6}
\end{figure*}

The central AGN is detected at 610 MHz with no radio emission 
associated with the intense star forming disk that is detected
in the NUV (Figure \ref{fig:figure.6} (a)).  Intense NUV 
emission indicates a recent burst of star formation
that has been triggered in the galaxy.  The south-western spiral arm shows a particularly
vigorous burst of star formation. Unresolved emission from the centre of the galaxy is 
detected  at all the three observed frequencies i.e. 1280~MHz, 610~MHz and 240~MHz 
(Figures \ref{fig:figure.6} (b)-(d)).  
The NUV emission is intense and defines the spiral arms of the galaxy whereas
the NIR emission is featureless with only an intense core being detected.  
A single power law fit to the three radio points resulted in a  spectral index
of -0.23$\pm$0.11 (Figure  \ref{fig:figure.9} (d)). 
However, the spectrum does not follow a simple power law as the spectral
index between 610 and 1280 MHz is -0.44$\pm$0.09 whereas between 240 and 610 MHz
is -0.05$\pm$0.01.

\textbf{UGC 6614}: This is a relatively well studied GLSB
galaxy with the AGN detected in mm-wave
continuum \citep{das06}, optical (\citealt{schombert98};
\citealt{ramya11}), X-ray \citep{naik10} and NIR \citep{rahman07} wavelengths.  
It has a low inclination and a large, prominent bulge
surrounded by a faint ring-like structure  \citep{rahman07}. This
ring is clearly seen in the Spitzer NIR image
(\citealt{hinz07}; \citealt{rahman07}) and simulations 
suggest that it is the result of galaxy collision in the 
past \citep{mapelli08}.  The radio spectrum of the core between 
100 GHz and 1.4 GHz is found to be flat \citep{das06}.  
It has been detected in CO emission. The emission is detected offset from
the galaxy centre and the molecular gas mass is 
$\sim$ 2.8 $\times$ 10$^{8}$ M$_{\odot}$ \citep{das06}. 
This galaxy is a member of the group LDCE 829 (Crook et al. 2007, 2008) 
which has three members.

\begin{figure*}
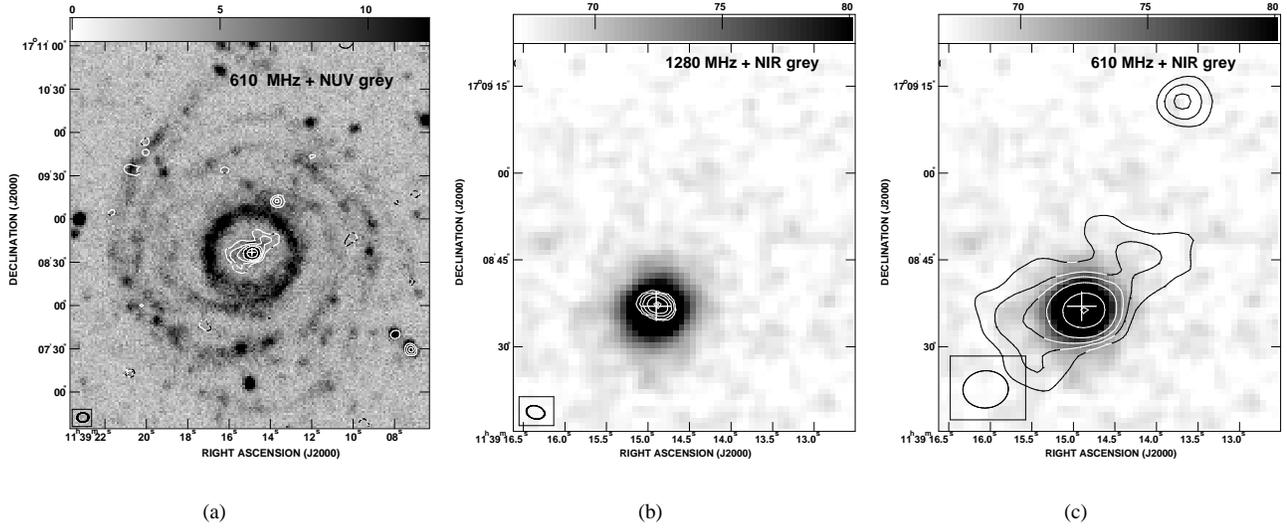

\subfigure[]{\includegraphics[height=7.0cm]{figure.7a.PS}}
\subfigure[]{\includegraphics[height=7.0cm]{figure.7b.PS}}
\subfigure[]{\includegraphics[height=7.0cm]{figure.7c.PS}}
\caption{\textbf {Images of UGC 6614}: The cross marks the optical centre of
the galaxy. 
(a) The contours showing the 610 MHz emission are plotted at  
0.1 $\times$ (-6,-4,4,6,8,12,24,39) mJy beam$^{-1}$. The 
angular resolution is 8$^{\prime\prime}$ $\times$ 6$^{\prime\prime}$,
PA = -85$^{\circ}$.92. The NUV grey scale is  counts/sec.
(b)  The contour levels of the 1280 MHz emission are  60 $\times$ 
(-8,-4,4,8,16,32,64,73) $\mu$Jy beam$^{-1}$ for beamsize  
3$^{\prime\prime}$ $\times$ 2$^{\prime\prime}$, PA = 74$^{\circ}$.26.
The NIR grey scale are data-number.
(c) Contour levels are similar to (a).}
\label{fig:figure.7}
\end{figure*}

Like Malin 2, NUV emission is detected over the entire disk 
with  star formation seen in the ring and along the spiral arms 
of the galaxy (Figure  \ref{fig:figure.7} (a)).  Radio 
emission at 610 MHz is mostly confined to the central region - from the 
AGN and extended along what appears to be jets/lobes along 
the north-west and south-east direction. No NUV emission
is coincident with these features.  Radio emission is detected
from several compact regions near the central ring
and along the spiral arms at 610 MHz. The emission is coincident 
with the NUV peaks indicating that its origin is star formation.    
Figures \ref{fig:figure.7} (b) and (c)  
show the zoomed-in contour images of the radio continuum 
emission from the galaxy at 1280~MHz 
and 610~MHz superposed on the 2MASS NIR image in grey scale.   
The AGN core is detected at 1280~MHz and NIR J band but no 
emission is detected from the radio jets/lobes.  
We detect a compact radio source to the north-west of the central core near right ascension 
11$^h$39$^m$13$^s$.75 and declination 17$^\circ$9$^\prime$13$^{\prime\prime}$ at 610 MHz  
coincident with NUV emission.
The estimated peak brightness of this source at 610 MHz is
0.9$\pm$0.2 mJy/b.   Using the $4 \sigma$ limit at 1280 MHz of 0.24 mJy implies
a spectral index steeper than $-1.7$ for this source.  This appears to be steep for a star
forming region in the galaxy and might be a background radio source which shows a
chance coincidence.  With the current data,
we are unable to comment on this region any further.

We estimate the peak flux density of 
the core to be  4.0$\pm$0.6 mJy at 610 MHz and the spectral index of 
the AGN core between 610 and 1280 MHz, 
$\alpha$ $\sim$ +0.12$\pm$0.05 (Table \ref{tab:table.5}).  We recall that the spectrum of
the central AGN between 111 GHz and 1.4 GHz was found to be flat by \cite{das06}.
The AGN is not detected in our 240 MHz map to a $4\sigma$ limit of 4.6 mJy 
and this is consistent with the estimated $\alpha$. 
We also estimated the spectral index of the lobe emission as follows.  
We subtracted the core flux density at 1280 MHz of 4.4 mJy from the
NVSS flux density of 7.5 mJy and assumed the remaining flux density as
arising in the jets/lobes.  In combination with the emission at
610 MHz obtained after subtracting the core emission from the total emission
at 610 MHz we obtain the  spectral index, $\alpha\sim -1.06\pm0.12$ between
610 and 1280 MHz for the jets/lobes. 

\textbf{UM 163}:  This galaxy, also known as 2327-0244, was
first studied by \cite{sprayberry95}  and  was found to host an AGN.  
The galaxy has a strong bar, 
a prominent bulge and two trailing spiral arms.  Its morphology
is similar to early type spirals but its disk is 
low surface brightness in nature. 

\begin{figure*}
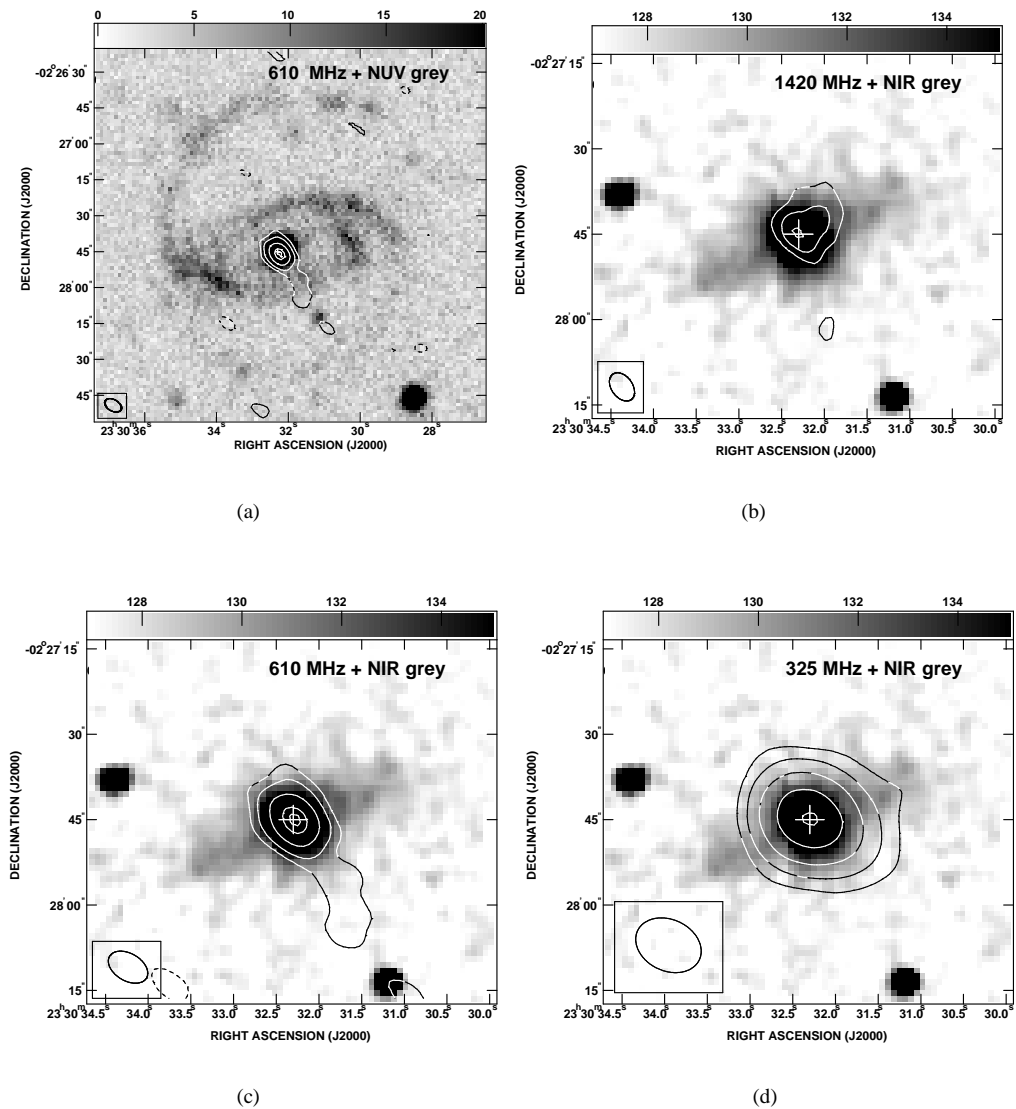

\subfigure[]{\includegraphics[height=7.0cm]{figure.8a.PS}}
\subfigure[]{\includegraphics[height=7.0cm]{figure.8b.PS}}\\
\subfigure[]{\includegraphics[height=7.0cm]{figure.8c.PS}}
\subfigure[]{\includegraphics[height=7.0cm]{figure.8d.PS}} \\
\caption{\textbf{Images of UM 163}: The cross marks the position
of the optical centre. 
(a) The contours showing the 610 MHz emission are plotted at 
0.2 $\times$ (-8,-4,4,8,16,25,28) mJy beam$^{-1}$. The angular
resolution is 7$^{\prime\prime}$ $\times$ 5$^{\prime\prime}$, PA = 59$^{\circ}$.41.  
The NUV grey scale is  counts/sec.
(b) The contour levels of the 1420 MHz emission are
0.2 $\times$ (-8,-4,4,8,14) mJy beam$^{-1}$ for beamsize 
5$^{\prime\prime}$ $\times$ 3$^{\prime\prime}$, PA = 36 $^{\circ}$.06. 
The NIR grey scale are data-number.
(c) Contour levels are similar to (a). 
(d) The contours showing the  325 MHz emission are plotted at 
0.4 $\times$ (-8,-4,4,8,16,32,50) mJy beam$^{-1}$.
The beamsize is 12$^{\prime\prime}$ $\times$ 9$^{\prime\prime}$, PA =64$^{\circ}$.70.}
\label{fig:figure.8}
\end{figure*}

This galaxy is detected at all the  observed frequencies.
NUV emission is detected from the centre of the galaxy; 
from a ring close to the centre and along the spiral arms 
(Figure \ref{fig:figure.8} (a)). The 610 MHz radio emission is
confined mainly to the core of the galaxy with a $4\sigma$ extension 
seen towards the south.  The radio emission is resolved at 
all the wavebands.  The bar is prominent in the NIR image 
(Figures \ref{fig:figure.8} (b)-(d)).  At 1420 MHz, 
the emission is extended towards the north-west. 
Our low resolution image at 325 MHz shows  the emission 
arising along the bar and also extended perpendicular to the bar. 
The emission is likely to be due to the AGN core and jets.  
The 610 MHz and the 1420 MHz maps are convolved to the 
resolution of the 325 MHz image to estimate
the spectrum of emission.   
We estimate  $\alpha$ $\sim$ -0.82$\pm$0.03 which  might indicate 
the dominance of lobe emission in the nuclear region. No emission at any radio 
frequencies  is associated with the spiral arms.

\section{Discussion}

\begin{table*}
 \centering
  \begin{minipage}{200mm}
\caption{The spectrum of the emission from the sample galaxies between 240 MHz 
and 1420 MHz along with nuclear flux densities due to AGN and disk emission}
\begin{tabular}{@{}llllllllllllll@{}}\hline

\footnotetext [1] {Robust =5 images convolved to lowest resolution to estimate integrated flux density.}
\footnotetext [2] {NVSS Flux density estimated from the NVSS images}
\footnotetext [3] {Likely to be due to the star forming disk and used in Table \ref{tab:table.6}}
\footnotetext [4] {The 325 MHz detects diffuse emission from the star forming disk which 
                    is not detected at the other frequencies, results in a steep spectrum of
                     $\alpha$=  -2.37$\pm$0.16.}
\footnotetext [5] {Jet/lobe emission included}

Galaxy   & Frequency & Nuclear Flux  &   $\alpha$ ($S \propto \nu^{\alpha}$) &   
       Integrated & $\alpha$  & NVSS Flux$^{b}$ & (NVSS - Nuclear  \\
&&Density&(Nucleus)&Flux Density$^{a}$ &(Integrated)&Density   & Flux Density)$^{c}$  \\
         &  MHz        &  mJy                  &                   &  mJy  & & mJy & mJy  \\ \hline 

UGC 1378 &  325 &  1.7$\pm$0.2               &                             &   2.2$\pm$0.3  & &       \\
         &  610 &  0.9$\pm$0.1               &  -0.9$\pm$0.1             &   1.9$\pm$0.2  & -0.29$\pm$0.03 &\\
         & 1420 &  0.54$\pm$0.02             &                             &   1.40$\pm$0.07&      
         &8.0$\pm$0.4 &7.4$\pm$0.4\\

UGC 1922 & 325  & 68.1$\pm$10.2             &                                  &  72.8$\pm$10.9 &&\\
         & 610  & 58.1$\pm$8.7              &   -0.39$\pm$0.09                 &   65.9$\pm$9.8 & -0.40$\pm$0.15& \\
         & 1420 & 36.5$\pm$1.8              &                                  &   37.8$\pm$1.8 & 
& 37.7$\pm$1.9&1.2$\pm$0.1\\  

UGC 2936 & 610  & 7.7$\pm$1.1 &  $\alpha^{610}_{1280}$ $\sim$  -0.44$\pm$0.09 &  48.0$\pm$7.2  
& $\alpha^{610}_{1280}$ $\sim$  -0.38$\pm$0.12 & \\
         & 1280 & 5.6$\pm$0.2   &  &  36.4$\pm$5.4 &            & 39.9$\pm$2.0   & 34.3$\pm$1.8\\

UGC 4422 &  325 & 9.5$\pm$1.4               &                                  &   49.9$\pm$7.4$^{d}$ & 
$\alpha^{325}_{610} \sim$    -2.37$\pm$0.16 &\\
         &  610 & 7.9$\pm$1.1               &  -0.41$\pm$0.07                  &   11.4$\pm$1.7 
& $\alpha^{610}_{1420} \sim$    -1.05$\pm$0.24  &\\
         & 1420 & 5.0$\pm$0.2               &                              &    4.6$\pm$0.2 &
&11.2$\pm$0.6 &6.2$\pm$0.4      \\

Malin-2  &  240 & 5.2$\pm$0.7   &                                          &   5.2$\pm$0.7  && \\
         &  610 & 5.0$\pm$0.7     &  -0.23$\pm$0.11                        &   5.0$\pm$0.7  & same as nuclear emission&\\
         & 1280 & 3.6$\pm$0.1  &                                           &   3.6$\pm$0.1  & 
& 6.9$\pm$0.3 & 3.3$\pm$0.2  \\

UGC 6614 &  240 &  $<$  4.6                 &                         &  $<$  6.8       &  & \\
         &  610 & 4.0$\pm$0.6 & $\alpha^{610}_{1280}\sim$  +0.12$\pm$0.05& 10.8$\pm$1.6 & 
$\alpha^{610}_{1280}\sim$  -1.13$\pm$0.17$^{e}$  &\\
         & 1280 & 4.4$\pm$0.2               &                                  &   4.4$\pm$0.2  & 
& 7.5$\pm$0.3 & 3.1$\pm$0.1   \\   

UM 163   &  325 & 26.1$\pm$3.9              &                                  &  27.9$\pm$4.1  &   &     \\
         &  610 & 15.2$\pm$2.2              &  -0.82$\pm$0.03                  &  15.6$\pm$2.3  & -0.77$\pm$0.11&\\
         & 1420 & 8.0$\pm$0.4               &                                  &   9.8$\pm$0.4  &
& 8.3$\pm$0.4 &0.3$\pm$0.0 \\
\hline
\end{tabular}
\end{minipage}
\label{tab:table.5}
\end{table*}

Our objective of the present investigation is to study the low radio frequency spectrum of the
central AGN, star forming disk and the influence of the environment
on our sample of GLSB galaxies using our high resolution, high sensitivity 
continuum images at  240, 325, 610 MHz  and  L band. 

\subsection{AGN in GLSB Galaxies}
\begin{figure*}
\vspace*{50pt}
\subfigure[]{\includegraphics[height=8.5cm,angle=-90]{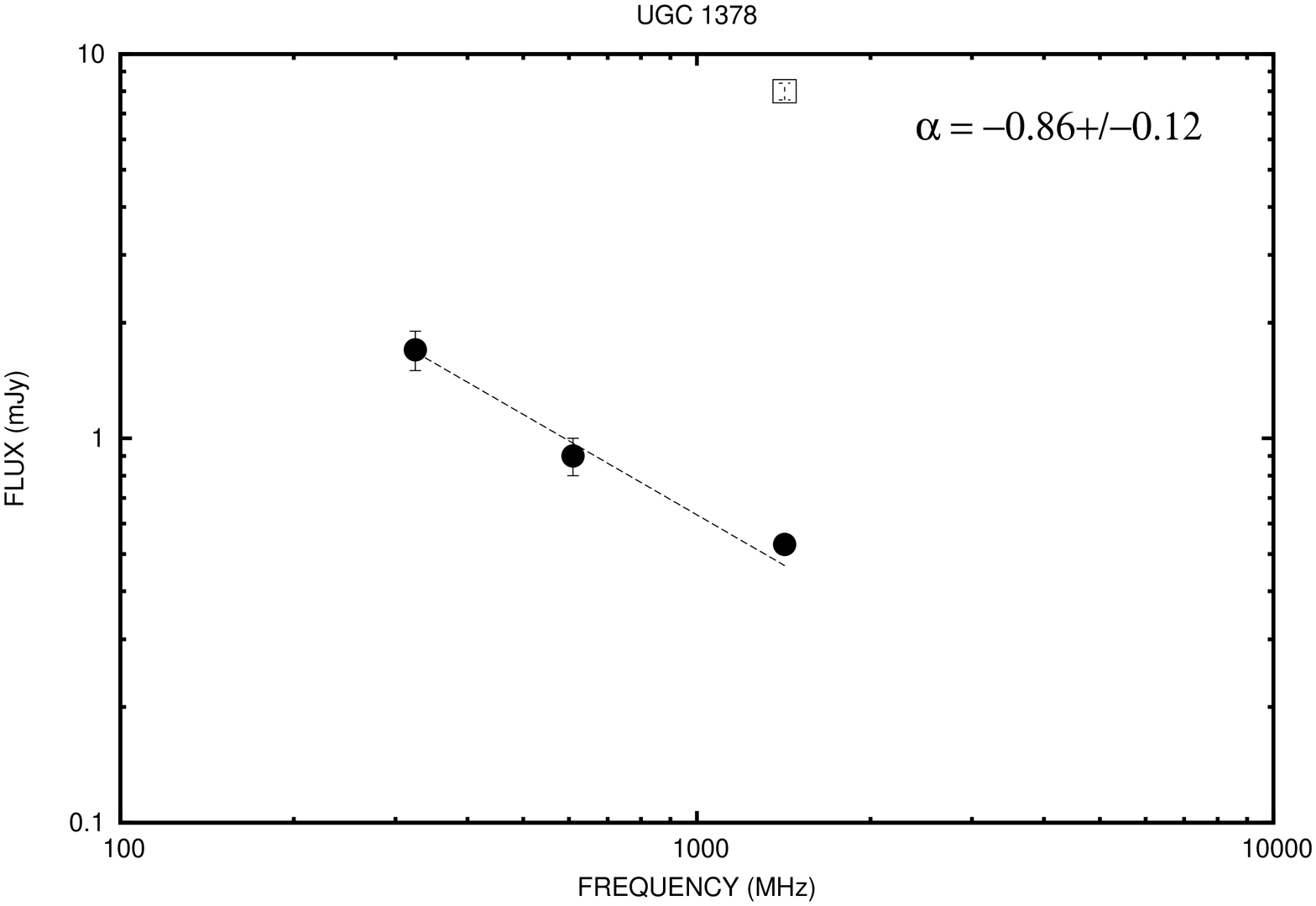}}
\subfigure[]{\includegraphics[height=8.5cm,angle=-90]{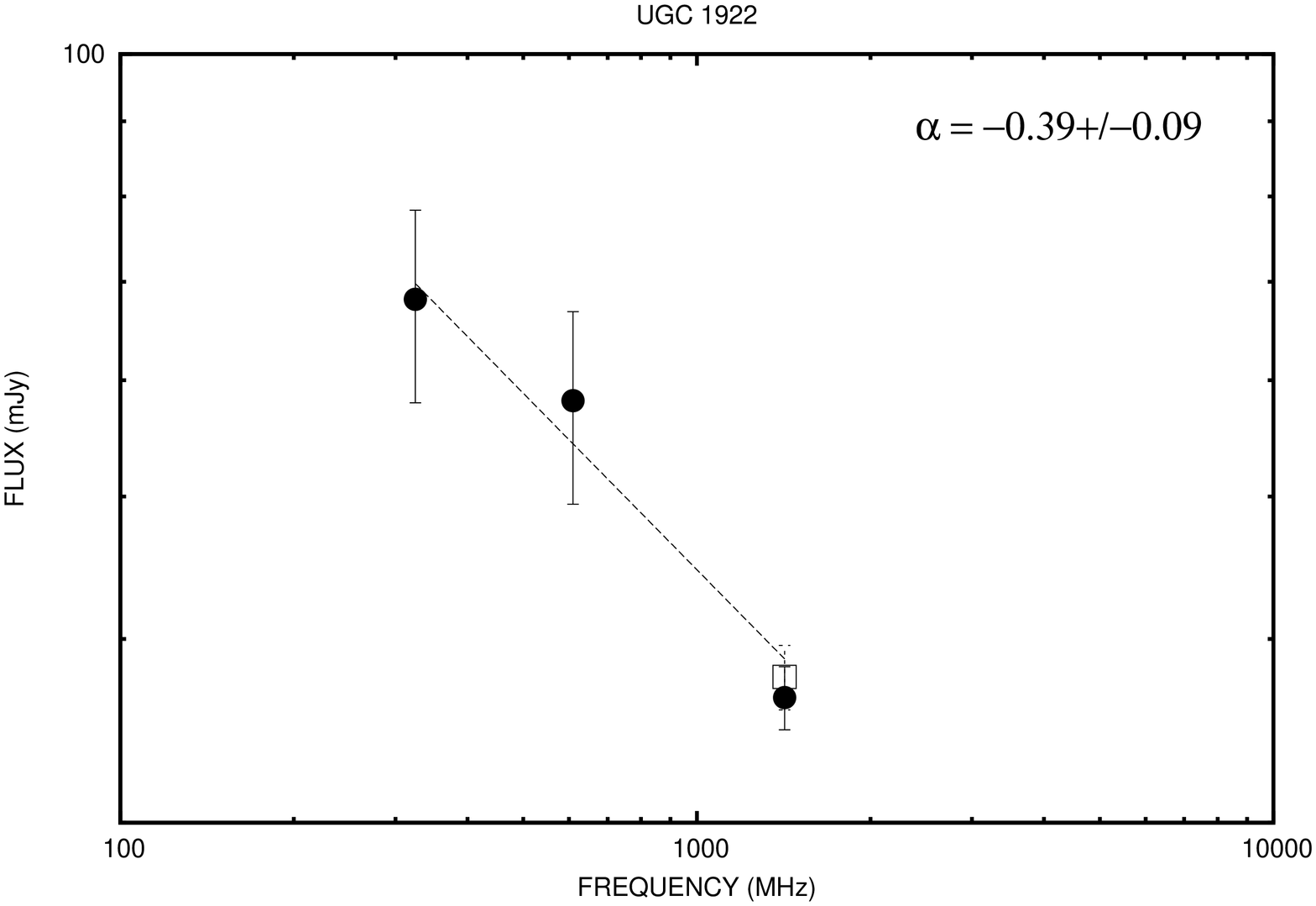}}\\
\subfigure[]{\includegraphics[height=8.5cm,angle=-90]{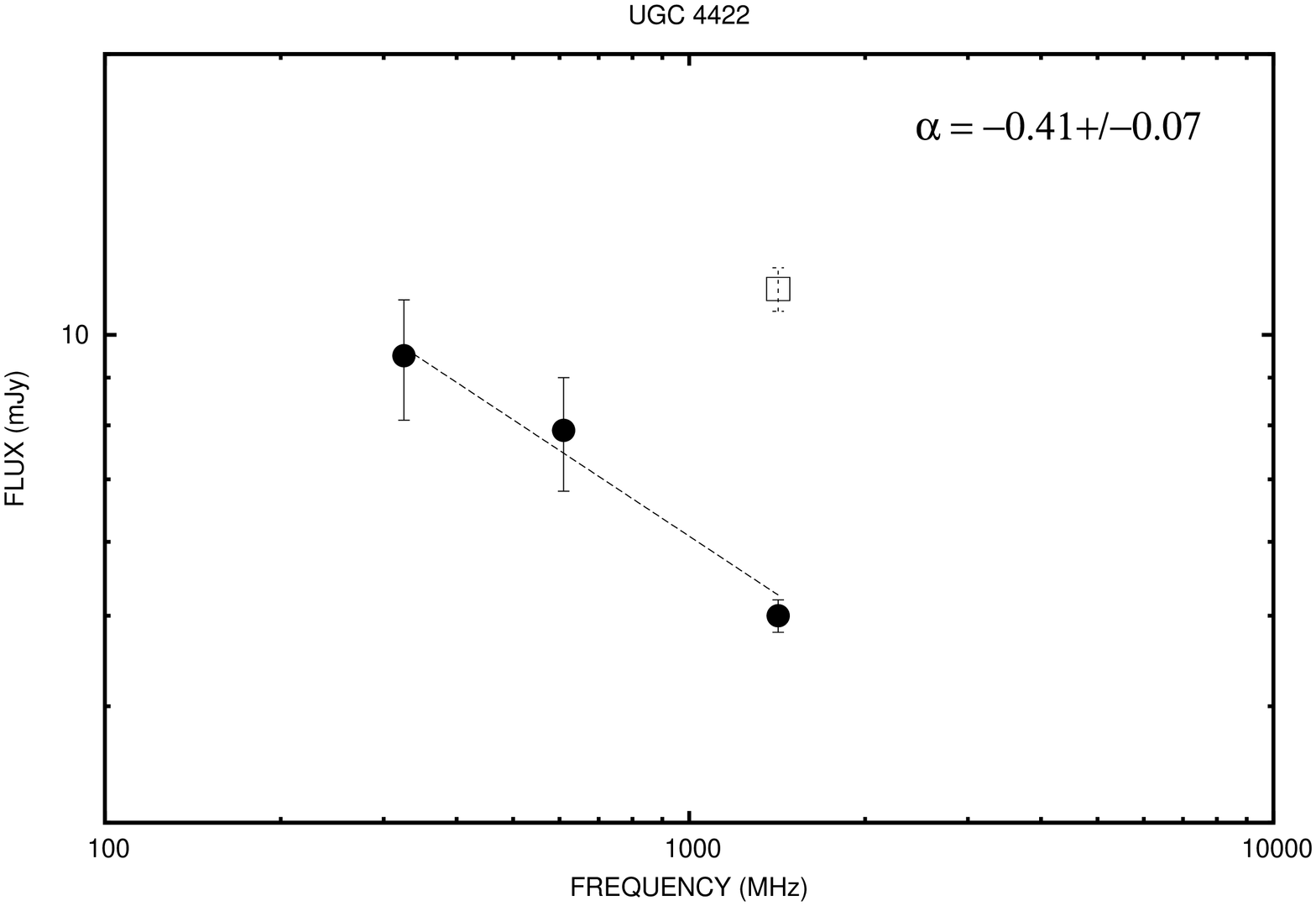}}
\subfigure[]{\includegraphics[height=8.5cm,angle=-90]{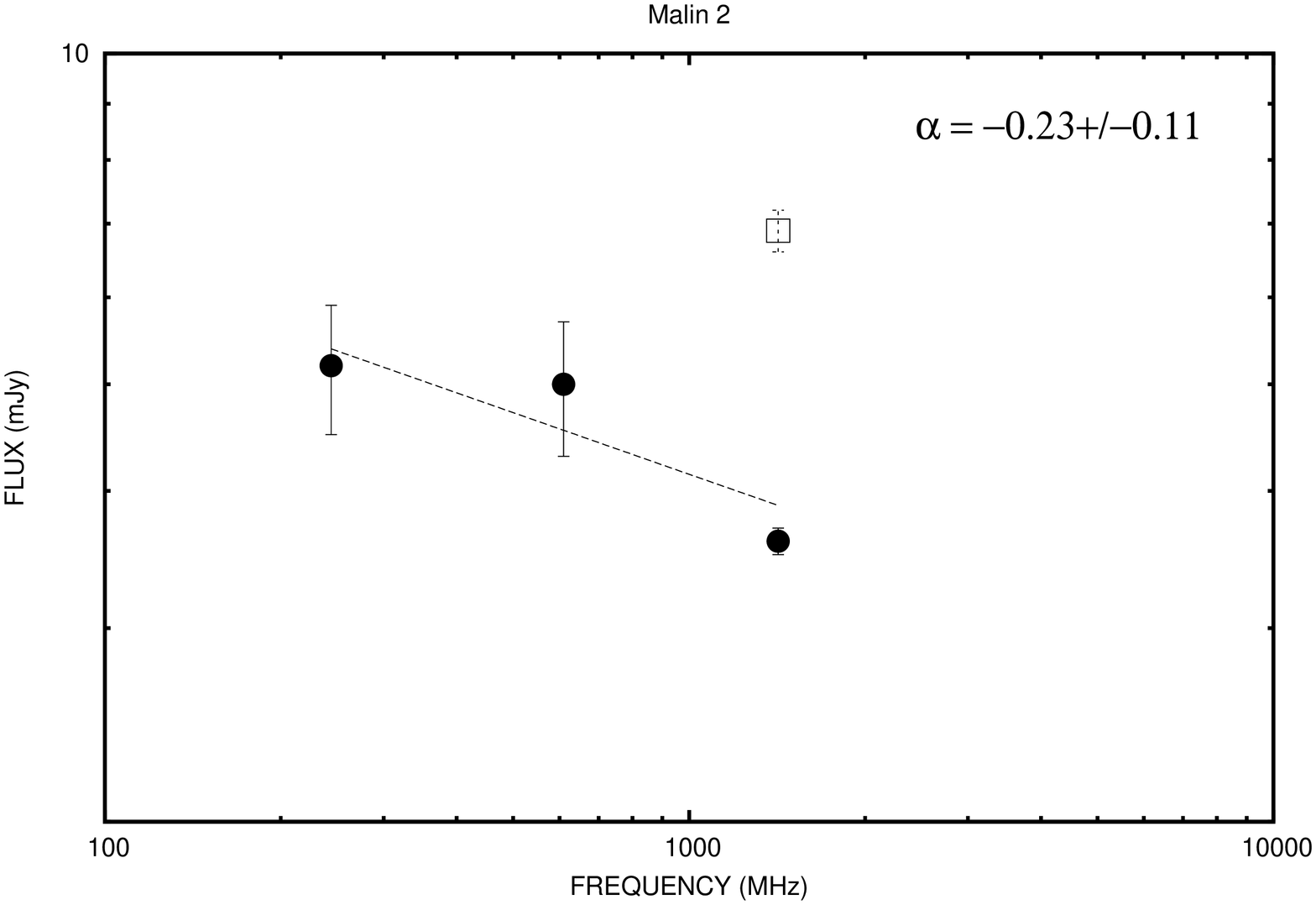}}\\
\subfigure[]{\includegraphics[height=8.5cm,angle=-90]{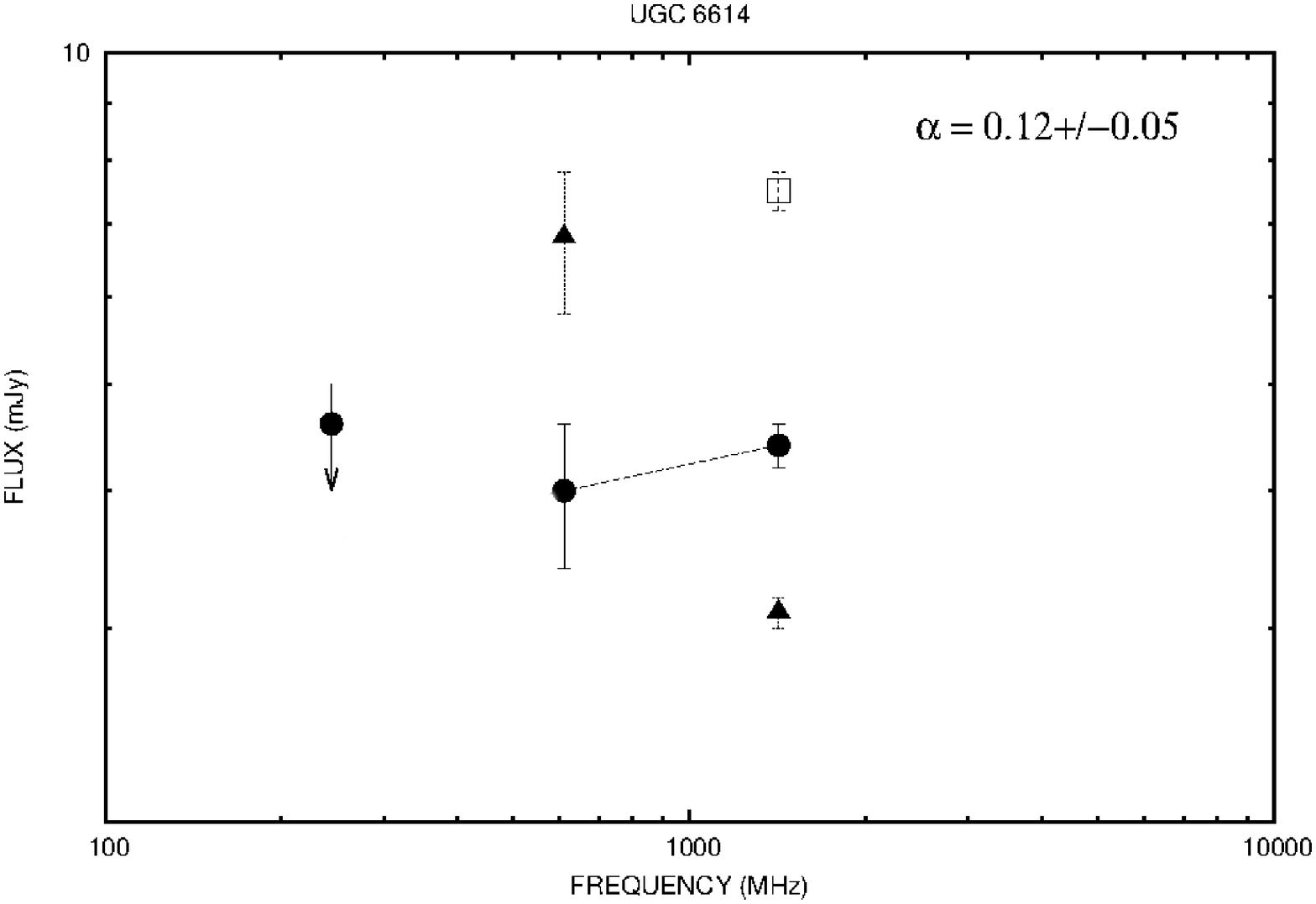}}
\subfigure[]{\includegraphics[height=8.5cm,angle=-90]{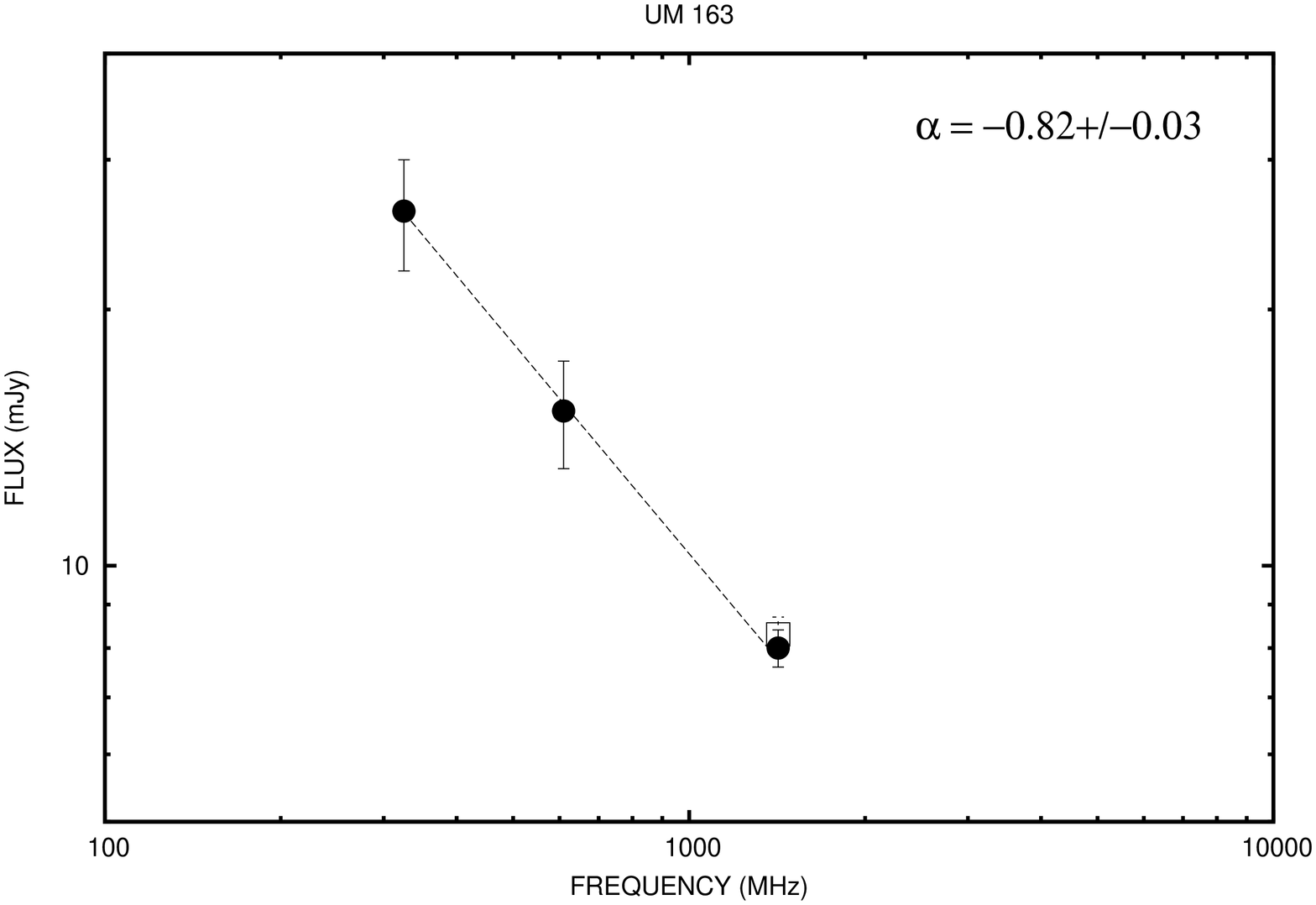}}\\
\caption{The spectral index $\alpha$ obtained from the fitted spectrum of 
the nuclear flux densities. Filled circles in the plots
show the nuclear fluxes. Open squares show the NVSS flux densities.
Filled triangle in the plot of UGC 6614 shows the jets/lobe flux density.}
\label{fig:figure.9}
\end{figure*}

 We detect radio continuum emission from the centre of all seven 
galaxies in our sample. Our high resolution data allows us
to determine the core  spectrum. We distinguish the spectrum  
as steep  if $\alpha$ $<$ -0.5 or flat if $\alpha$ $\geq$ -0.5; 
$S \propto \nu^{\alpha}$. In AGNs, compact 
flat-spectrum nuclear radio cores are widely accepted as 
the signature of  synchrotron self-absorption.   
The radio spectra of the nuclear sources of the galaxies, 
UGC 1922, UGC 2936, UGC 4422, Malin 2 and UGC 6614 exhibit  
spectral indices ranging from  $\alpha = +0.12$ to $-0.44$.
We interpret these  flatter spectra as arising 
from the radio core. In case of UGC 1378 and UM 163 the 
spectrum is steeper. 
We note that in UGC 1378, the peak
radio emission is displaced to the south of the optical centre indicating
an origin distinct from the AGN.  Only faint radio emission is
detected from the optical centre which is likely to be due to
the radio core.  However, we are not able to estimate the spectral index of 
this source separately. In case of UM 163, it is likely
that the radio emission includes contribution from the jets/lobes
thus leading to its steep spectrum.  We note that the 
flux density measured by NVSS is similar to our high resolution 
L band image (see  Figure \ref{fig:figure.9} (f)).

Recent  studies
show that only about 15 - 20\% of normal LSB galaxies host
an AGN  (\citealt{burkholder01}, \citealt{mei09}) and
most appear to be associated with large bulges \citep{schombert98}.
The nuclear black hole masses of GLSB galaxies are relatively low and
lie in the range $10^{5} - 10^{6}~M_{\odot}$ \citep{ramya11}.
GLSB galaxies lie below the M$_{BH}-\sigma$ relation followed by other galaxies
\citep{ramya11}.   Thus though these galaxies are very massive,
their AGN  are less evolved than brighter galaxies on
the $M-\sigma$ correlation.
This is possibly because the dark matter halos in these 
galaxies inhibit the formation of disk instabilities that 
trigger star formation and lead to gas infall into their 
nuclear regions \citep{mayer04}. The slower  gas infall rate 
probably results in a lower AGN fueling rate, which leads 
to lower nuclear black hole masses.

Thus, on the basis of an analysis of low frequency radio data 
of a sample of seven GLSB galaxies with optically
identified AGN, we find that five out of them have radio emission  
associated with AGN core having  spectral indices $\geq$ -0.44,
thereby exhibiting signature of  synchrotron self-absorption.   
Two  galaxies of our sample, UGC 6614 and UM 163, appear to have  
additional emission from jets/lobes.  
However, higher resolution images are required 
to disentangle the contributions from the core of the 
AGN and the jets/lobes in these two cases. 

\subsection{AGN-Jets in GLSB Galaxies} 
Radio jets are often considered 
to be tracers of AGN activity in the more massive radio galaxies
even if the core is not detected. 
However, radio jets in spiral galaxies are relatively rare 
probably  because their AGN are relatively weaker than those 
found in the more massive radio galaxies. 
\cite{gallimore06} studied a sample of 43 Seyfert galaxies 
and found that 19 (44 $\%$) showed extended radio jets.  In our sample
of seven GLSBs, we clearly detect a jet in UGC 6614 and a jet-like feature
is detected in UM 163.   Thus we detect jet-like features in 2 out of
7 galaxies i.e. about 28\%.   However the sample size needs to be increased to
make a valid comparison.
The Seyfert nucleus of the spiral NGC~4258 is one of rare examples of extended 
jets in spiral galaxies \citep{krause07} as is NGC 1275 (Perseus A). 
The jet length scales in UGC 6614 and UM 163 are 6.8~kpc 
and 13~kpc respectively and both jets end 
well within the inner optical disk, which is similar 
to what is seen in other spirals \citep{laine08}.  The spectrum of emission 
from the jets in UGC 6614 has a spectral index of -1.06$\pm$0.12. 
Thus the radio properties of AGN in GLSB galaxies appear to be similar 
to that observed in brighter spiral galaxies.

\subsection{Comparison With Other Wavebands}
All the seven galaxies in our sample have GALEX UV and 
2MASS NIR images.  Surprisingly, five  of the 
galaxies, UGC 1922,  UGC 4422, Malin 2, UGC 6614 and UM 163 
show bright NUV emission over their entire disk which could be 
due to a recent burst of massive star formation in the galaxy. 
All the galaxies show radio emission 
associated with the nuclear region, additionally UGC 2936 
and UGC 4422  show radio emission associated 
with the star forming disk.
The NVSS flux densities at 1.4 GHz 
are higher than what we record in our high 
resolution L band images for many of the galaxies 
(see Table \ref{tab:table.5}).  We infer that the excess radio
emission in most cases is from the disk of the galaxies.  
In particular, we find that more than half of the emission 
recorded by NVSS at 1.4 GHz is associated with a diffuse, 
disk component for UGC 4422, UGC 1378 and UGC 2936 
(Table \ref{tab:table.5}).  On the other hand,  80\% of
the total emission at 1.4 GHz from  
UGC 1922 and UM 163  and  50 \% emission from Malin 2 and 
UGC 6614 arise in the active nucleus. 
All these galaxies show bright extended NUV disks indicating  a 
recent burst ( $\sim$ 100 million yrs)  of star formation.  
This starburst is yet to result in sufficient number of supernova
remnants required for the non-thermal radio emission to be detectable.
The synchrotron lifetime is  expected to be 
100 Myr at 1.5 GHz if strength of magnetic field (B) is 
5 $\mu$G \citep{condon92}.
Moreover in case of UGC 4422,  NUV, NIR and radio emission 
are all detected from the disk indicating  continuous 
star formation in the galaxy. It would be interesting to 
derive estimates of the stellar ages in these
galaxies.

We detect diffuse radio emission from the disk of 
UGC 2936.  Faint NUV is detected from the centre of the  
galaxy.  The presence of NIR
emission and absence of NUV  from the disk of the galaxy suggests
an older burst of star formation, probably more than few Gyr ago. 
The star formation episode
has since been quenched and the galaxy is  not prominent in the NUV. 
No extended radio emission is detected from the disks of 
UGC 1922, Malin 2, UGC 6614 and UM 163.

The combination of our radio results with the NIR and NUV results
seem to give support to the episodic star formation in our sample GLSB galaxies.  
It is not clear what leads to a starburst episode or what leads to it
being quenched but the range of features displayed by our sample of
GLSB galaxies suggests that these galaxies are intriguing  
and might likely be an evolutionary step for many 'normal' galaxies.
GLSB evolve  at a much slower rate because of the effect of the dominant 
dark matter halos.

\subsection{Star Formation Rate in GLSB Galaxies} 
LSB  galaxies generally have a lower SFR 
($\sim 0.2$ M$\odot$/yr; \citealt{mcgaugh94})  
and lower metallicity than HSBGs. 
High resolution imaging of star forming 
regions in LSB  galaxies show 
that their disks contain HII regions that are
very similar to those found in brighter galaxies 
\citep{schombert13}. 
Their disks are gas rich but 
the surface density is below the critical threshold for triggering 
star formation (\citealt{vanderhulst93}; \citealt{pickering97}; \citealt{das10}). 
We estimated the SFR in the sample of our galaxies using the 
relation as given in \cite{condon92} for M $\geq$5 M$\odot$  and
the results are reported in  Table \ref{tab:table.6}.  
For this, we subtracted the flux density of the nuclear source as obtained  
from our high resolution images at L band from the flux density 
of the galaxy, estimated from the lower resolution NVSS image. 
We find that for UGC 1922 and UGC 2936, the flux densities 
estimated from our low resolution GMRT maps are similar to the NVSS values.
In UM 163, our low resolution GMRT map gives a larger flux density than 
the NVSS map and for the remaining galaxies the NVSS flux densities are larger.
 We assume  that the excess NVSS emission is due to  disk star formation, 
and non-thermal in nature. 
Using above flux densities and assuming the non-thermal spectral 
index $\alpha = -0.8$,  we estimated  average star  formation 
rates listed in  Table \ref{tab:table.6}. 
The radio emission can be used as a direct probe 
of  the very recent star-forming activity in normal and starburst 
galaxies. Nearly all of the radio emission at lower frequencies  
from such galaxies is synchrotron radiation from  
relativistic electrons and free-free emission from H II regions. 
Massive stars (M $\geq$5 M$\odot$) produce the supernovae whose 
remnants (SNRs) accelerate  the relativistic electrons 
and are  used as a tracer of  star  formation.
It may be noted from  Table \ref{tab:table.6} that the SFRs of the sample galaxies  range from 
0.15 to 3.6 M$\odot$/yr with five galaxies showing SFRs less than 
1 M$\odot$/yr.
For comparison, the SFRs that we 
obtained from literature which were estimated using either the UV or
total infrared emission are also listed there.  These range from 0.8 to 4.3
M$\odot$/yr.   This suggests that the SFRs of these galaxies show a wide
range similar to the HSBGs and are likely for different epochs.    

\cite{boissier08} have examined the UV properties of a sample of 
13 GLSB  galaxies which include three galaxies from our 
sample; UGC 2936, Malin 2 and UM 163. Using GALEX data they 
find that the UV disks of 9 out of their sample of 
13 galaxies are extended and the FUV-NUV colours of these galaxies
are redder than normal galaxies which they interpret as being due to episodic 
bursts of star formation.  Four galaxies in our sample  show bright 
NUV disk.

We  estimated supernova rates which range from 0.006 to 0.15 yr$^{-1}$ (see Table \ref{tab:table.6}). 
For comparison the supernova rate in M82 is  $ 0.1 yr^{-1}$ \citep{condon92}. 
Two of the galaxies in our sample, namely UGC 1922 and UGC 4422 have
hosted a type Ia supernova - SN1989s and SN1999aa respectively.  Since the
progenitor system of these type of supernovae are believed to consist of a binary
with a white dwarf member, it can indicate star formation which was
triggered more than a few Gyrs ago in these galaxies with the low mass stars evolving
into type Ia supernovae. We note that NUV, NIR and radio continuum emission from
the stellar disk are detected in the case of UGC 4422 which could indicate
continuous star formation. In case of UGC 1922, most of the
resolved radio continuum arises close to the centre of the galaxy.  NUV is detected
from arcs resembling spiral arms around the centre of the galaxy whereas
diffuse NIR is detected from the centre of the galaxy and from a couple of massive star
forming regions to the north-east of the centre of the galaxy.
On the other hand, UGC 2936 has hosted a core-collapse
type II supernova (SN 1991bd). We note that while the star forming disk
of UGC 2936 is detected in the radio and NIR bands, only faint NUV emission is detected
from the centre of this galaxy (see Figure \ref{fig:figure.3} (a)).  
This could indicate an older star forming episode, say a few Gyrs ago which
would be responsible for the NIR emission and a more recent star forming episode which would
be responsible for the radio emission and the detected supernova type II.  Lack of 
NUV emission could indicate that the later star formation episode might have since been quenched.  
We note that it might be possible to verify these results by a detailed study of the age of the
stellar populations in these galaxies which is beyond the scope of this paper.
Group membership is confirmed for UGC 1922 and UGC 4422 whereas we could not find
any information for UGC 2936 in literature.   While this is a qualitative picture,
a more quantitative picture of the multi-band emission might be able to
result in a more complete picture of the timescales that the different 
diagnostics tracers.

Another measure of the AGN versus starburst nature 
of the radio emission is the FIR-radio correlation parameter q. 
The normal galaxies have a value of q= 2.35 (\citealt{condon92}; 
\citealt{yun01}).  We also estimated FIR-radio (1.4 GHz) 
correlation parameter q for our sample of galaxies. 
The FIR data for four of the sample galaxies, available from  
IRAS Faint Source Catalog (NED) are  listed in column 7 of Table \ref{tab:table.6}.  
We find the galaxies UGC 1378, UGC 2936, UGC 4422 and UM 163 in 
our sample follow the FIR-radio correlations although UM 163 
has low q value (see Table \ref{tab:table.6}).

\subsection{The Environment of GLSB Galaxies and Star Formation}
We examined the environment of the sample galaxies. 
Four of the sample galaxies are reported in  literature to have 
group membership;  UGC 6614 (LDCE 829), UGC 4422 (LGG 159, LDCE 571), 
UM 163 (LDCE 1583), UGC 1922 (LDCE 163). LDCE (Low Density Contrast Extended) 
groups have been catalogued by Crook et al. (2007, 2008) from 
2MASS with the number of members ranging from 3 to 37. 

A study of small scale environment of GLSB galaxies  \citep{bothun93} has 
revealed that there is a deficit of galaxies 
located within 0.5 Mpc and within a velocity of 500 kms$^{-1}$. 
The distance 
to the nearest neighbour for GLSB galaxies is about 1.7 times 
farther than  for HSBGs  \citep{bothun93}.
Contrary to this, \cite{sprayberry95} inferred from their study that the 
small scale environment of GLSB galaxies was similar to that of 
HSBGs. Thus, it is not clear if the  unevolved LSB  
disks are due to the lack of nearby companions which would tidally 
trigger star formation.

UGC 4422, which is catalogued as a member of LGG 159 (Garcia 1993) 
is  the only galaxy in our sample whose disk is bright in radio, 
NIR and NUV and hence is likely to be subjected to enhanced and 
possibly continuous tidal interactions  which  play an
important role in the evolution of the  galaxy.  
All the four galaxies which are in groups show
an extended UV disk in the GALEX data indicating a recent 
burst of star formation.  The  tidal interaction in the 
group environs is likely responsible for this.
Interestingly, UGC 1378 and UGC 2936, which are  known not to have any group membership, 
the integrated radio emission appears to include large contribution from
a star forming disk. In  UGC 2936 the observed faint NUV  emission  is 
interpreted as  indicating  
an older episode of star formation.  The only isolated GLSB galaxy
which shows bright NUV emission is Malin 2 and the source of  the trigger of star formation in
this galaxy is likely internal.  We note that, the discussion here supports
the episodic star formation in GLSB galaxies  with bursts of star formation followed
by a quiescent phase as suggested by \cite{boissier08} from their GALEX UV study of 
13 LSB galaxies.

\section{Conclusions}

\begin{table*}
 \centering
  \begin{minipage}{100mm}
   \caption{Star Formation Rate and FIR-Radio Correlation (q) for Galaxies}
\begin{tabular}{@{}lllllllllll@{}}\hline
\footnotetext[1]{SFR$_{NUV}$ taken from   \cite{boissier08}}
\footnotetext[2]{SFR$_{TIR}$ estimated from the total IR luminosity (L$_{TIR}$)
using the relation between L$_{TIR}$ and SFR$_{TIR}$  \citep{bell03}}
\footnotetext[3]{SFR$_{TIR}$ taken from   \cite{rahman07}}
\footnotetext[4] {L$_{1.4GHz}$ and SFR$_{1.4GHz}$ estimated from the difference between NVSS
                 flux density and GMRT nuclear flux density listed in Table \ref{tab:table.5}}
\footnotetext[5]{ Supernova rate and SFR$_{1.4GHz}$ estimated using the relation given in \cite{condon92}}

Galaxies & SFR$_{NUV}$ $^{a}$ &SFR$_{TIR}$ & L$_{1.4GHz}$$^{d}$ &  SFR$_{1.4GHz}$$^{d}$ &  SNR$^{e}$ & logFIR & q  \\
& M$_{\odot}$ yr$^{-1}$& M$_{\odot}$ yr$^{-1}$ & 10$^{21}$ (WHz$^{-1}$) &  M$_{\odot}$ yr$^{-1}$
& yr$^{-1}$ &10$^{-14}$(Wm$^{-2}$)& \\ \hline

UGC 1378 & ....  &   ....     & 1.2  &  0.28 &  0.01 & 6.19     &  2.3   \\
UGC 1922 &  .... &   ....     & 3.1  &  0.75 &  0.03 & ....     &  ....  \\
UGC 2936 & 0.84  &   ....     & 10.0 &  2.45 &  0.12 & 28.42    &  2.2   \\
UGC 4422 & ....  & 2.80$^{b}$ & 2.9  &  0.70 &  0.03 & 6.68     &  2.2   \\
Malin-2  & 4.30  &  ....      & 15.0 &  3.60 &  0.15 & ....     &  ....  \\
UGC 6614 & ....  & 0.88$^{c}$ & 3.0  &  0.74 &  0.03 & ....     &  ....   \\
UM 163   & 3.30  &   ....     & 0.6  &  0.15 & 0.006 & 4.00     &  2.1    \\
 \hline
\end{tabular}
\end{minipage}
\label{tab:table.6}
\end{table*}

We have mapped the radio continuum emission at 
L band, 610, 325 and 240~MHz of a sample of 
seven GLSB galaxies using the GMRT. 
All the galaxies host an optically identified AGN.  Below we give the
summary of our results and list our main conclusions.

{\bf 1.~}We detect compact radio emission from the centres of
all the sample GLSB galaxies. The spectra of five galaxies
(UGC 1922, UGC 2936, UGC 4422, Malin2 and UGC 6614) have spectral indices 
ranging from 0.12 to -0.44 and that of  the galaxies UGC 1378 and UM 163 
exhibit a steeper spectrum.  Two of the galaxies UGC~6614 and UM~163 show
extended emission associated with their nuclei but show no correlation
with star formation traced by other diagnostics.  We interpret
the extended emission as being due to the radio jets or lobes of 
the active nucleus.  In UGC~6614 the radio jet extends out 
to a radius of 6.8~kpc and in UM~163 it extends out 
to 13~kpc.  In both cases, the jet lies within the optical disks.
Radio jets are relatively rare in spirals.

{\bf 2.~} Diffuse radio continuum emission associated with star 
formation in the disk is detected from the galaxies UGC 2936 and UGC 4422
at one of our observed frequencies. We used our high resolution maps at 
L band and the NVSS maps at 45$^{\prime\prime}$ resolution to
separate  the nuclear emission and disk emission.  The radio emission outside the
nucleus was presumed as being due to star formation and SFRs were found to
range from  0.15 to 3.6 M$\odot$/yr.  

{\bf 3.~} All the galaxies in our sample have been observed in the UV by GALEX
and in the NIR by 2MASS thus allowing us to make a multi-wavelength study of
this sample.  All our sample galaxies have bulges that are prominent in NIR.
Extended UV disks are detected in five galaxies namely UGC 1922, UGC 4422, Malin 2,
UGC 6614 and UM 163. NIR  disk/bar emission is detected  from UGC 2936, 
UGC 4422 and UGC 1378.  UGC 4422 is the only sample galaxy which shows extended 
UV disks and NIR  disk/bar emission. We suggest that a recent burst of star 
formation has occurred in the five galaxies with extended UV disks out of which
we find four reside in group environments.   
We find that most of the radio emission
in UGC 1922 and  UM 163 arises in the active nucleus.  Since these
two galaxies show bright NUV emission,  this suggests a fresh star 
forming episode.  The massive stars are yet to evolve
into supernovae and give rise to non-thermal emission. 
On the other hand, hardly any NUV emission is detected from  
UGC 2936 but the galaxy shows that more than 80\% of 
its L band radio emission arises from its star forming disk. 
This argues for  a star formation episode
which is more than 10-100 million years
old and has since been quenched.  It would be interesting to 
estimate stellar ages in these galaxies to confirm this scenario. 

{\bf 4.~} The study of these seven galaxies suggests that the environment in
which the galaxies are evolving is an important trigger for star formation.
It would be interesting to extend this study and examine the environment of 
a large sample of GLSB galaxies which is now possible with several catalogue
of groups of galaxies and of isolated galaxies.
Finally, we speculate that the low surface brightness phase 
of evolution might be an important evolutionary step for 
most of the disk galaxies, especially for galaxies evolving 
in poor environment.

\section*{Acknowledgments}
We feel highly obliged to the referee for  
constructive comments  which has improved 
the original manuscript.   
We thank the staff of the GMRT who made the observations 
possible. The GMRT is operated by the National Centre for 
Radio Astrophysics (NCRA), Pune of the Tata Institute of Fundamental 
Research. AM would  like to thank the NCRA-TIFR, Pune and  
Indian Institute of
Astrophysics, Bangalore for hospitality. 
This research has made use of NASA/IPAC Infrared
Science Archive,  the NASA/IPAC Extragalactic Data 
base (NED),  GALEX and 2MASS which is a NASA mission 
are operated by Jet Propulsion Laboratory, California Institute of 
Technology under contract with the National
Aeronautics and Space Administration.  We also like 
to acknowledge the site  http://skyview.gsfc.nasa.gov/.

\bibliographystyle{mn2e}
\bibliography{lsb-ms11may14}

\end{document}